\documentclass[%
 reprint,
 amsmath,amssymb,
 aps,
]{revtex4-1}

\usepackage{graphicx}
\usepackage{dcolumn}
\usepackage{bm}
\usepackage{color}
\usepackage{xspace}

\begin{document}

\preprint{APS/123-QED}

\title{The role of phase dynamics in a stochastic model of a passively advected scalar}
\author{Sara Moradi$^1$}
\email{smoradi@ulb.ac.be}
\author{Johan Anderson$^2$}%

\affiliation{%
$^1$ Fluid and Plasma Dynamics, Universit\'{e} Libre de Bruxelles, 1050-Brussels, Belgium\\
 $^2$ Department of Earth and Space Sciences, Chalmers University of Technology, SE-412 96 G\"{o}teborg, Sweden}

\begin{abstract}
Collective synchronous motion of the phases is introduced in a model for the stochastic passive advection-diffusion of a scalar with external forcing. The model for the phase coupling dynamics follows the well known Kuramoto model paradigm of limit-cycle oscillators. The natural frequencies in the Kuramoto model are assumed to obey a given scale dependence through a dispersion relation of the drift-wave form $-\beta\frac{k}{1+k^2}$, where $\beta$ is a constant representing the typical strength of the gradient. The present aim is to study the importance of collective phase dynamics on the characteristic time evolution of the fluctuation energy and the formation of coherent structures. Our results show that the assumption of a fully stochastic phase state of turbulence is more relevant for high values of $\beta$, where we find that the energy spectrum follows a $k^{-7/2}$ scaling. Whereas for lower $\beta$ there is a significant difference between a-synchronised and synchronised phase states, and one could expect the formation of coherent modulations in the latter case.     
\end{abstract}

\pacs{Valid PACS appear here}
\maketitle

\section{Introduction} 
In turbulence theory the phase dynamics of the fluctuations have hitherto been largely neglected and most often random phases are assumed. One would think that the rate of stochastization occurring in non-linear interactions increases as the amplitude of the modes increase, however the increased energy flux through non-linear mode coupling processes allows for coherent structures (sheared flows, geodesic acoustic modes etc. \cite{hillesheim2016}) to be formed which can even grow in the presence of random fields or cascade to smaller scales. This is one manifestation of the so-called self-organization process. The self-organisation phenomena is common in the everyday world and is a rapidly developing area of research within the field of dynamical chaos theory. There are a number of interesting examples that all contribute to this growing field such as biological clocks, physiological organisms and chemical reactors, just to name a few \cite{winfree, kuramoto, daido1, daido2, crawford, daido3, strogatz, hong, sonnenschein, kim2003}. A powerful and yet simple mathematical framework for describing the self-organisation phenomena was developed by Kuramoto \cite{kuramoto, Acebron}. The Kuramoto model describes the phase dynamics of a system of stochastic limit-cycle oscillators running at arbitrary intrinsic frequencies, and coupled through the sine of their phase differences, while under certain conditions they spontaneously lock into a common frequency. This paradigm of phase synchronization covers a diversity of physical situations such as those named above. 

In plasma turbulence theory however, due to the complexity of the system with many non-linearly interacting waves, the dynamics of the phases is often disregarded and the so-called random-phase approximation (RPA) is used assuming the existence of a Chirikov-like criterion for the onset of wave stochasticity \cite{Zaslavski1967, Zakharov1984}. In this approximation one assumes that the dynamical amplitudes can be represented as complex numbers, $\psi=\psi_r+i\psi_i=ae^{i\theta}$, with the amplitudes slowly varying whereas the phases are rapidly varying and, in particular, distributed uniformly over the interval $[0;2\pi)$. However, one could expect that the phase dynamics can play a role in the self-organisation and the formation of coherent structures as was shown in ref. \cite{GuoPRL2015, GuoPoP2015}. In the same manner it is also expected that the RPA falls short to take coherent interaction between phases into account. Moreover along the same lines, a model of stochastic oscillators obeying predator-prey rate equations have been developed to study the coupled dynamics of drift wave - zonal flow (DW-ZF) turbulence by which the system regulates and organizes itself. Within each population of DW-ZF a Kuramoto-type competition between the phases was assumed with an additional linear cross-coupling between the dual populations. Thus, the synchronization state of the whole system is controlled by two types of competition. It was shown that the system undergoes a modulational synchrony transfer between the two populations similar to the predator-prey oscillations in DW-ZF system \cite{moradi2Dmodel}.

In this paper therefore, we aim to study the role of phase dynamics and the coupling of phases between different modes on the characteristic time evolution of the turbulent fluctuation and the formation of coherent structures. In order to elucidate on the phase dynamics we assume a simple turbulent system where the so-called stochastic oscillator model can be employed. The idea of interpreting turbulence by stochastic oscillators goes back to Kraichnan \cite{Kraichnan1961}, where a novel approach was offered to capture several important features of the turbulent dynamics. The stochastic oscillator models can be derived from radical simplifications of the nonlinear terms in the Navier-Stokes or Gyro-Kinetic equations. In this particular case we adopt the basic equation for the stochastic oscillator model with passive advection and random forcing from Ref. \cite{krommes2000b}: 
\begin{eqnarray}
\partial_t \psi + u(t).\nabla \psi = \hat{f}^{ext}(t),
\label{eq0}
\end{eqnarray} 
where $u(t)$ and $\hat{f}^{ext}(t)$ are random values with given statistical properties depending on the problem of interest. The eq. (\ref{eq0}) is a linear equation in scalar $\psi$, however  since $\psi$ and $u$ are considered to be random fluctuating quantities, the second term on the left hand side is quadratically non-linear in random variables. The model is intended to capture several important features of the typical quadratically nonlinear primitive equations that arise in practice. Here, we introduced a non-linear phase coupling dynamic described by an extended Kuramoto type equations previously described in \cite{moradi2Dmodel} between the two random quantities namely passive advective flow and the forcing. In our model the non-linear amplitude coupling is replaced by a prescribed dispersion relation for the natural frequencies of the phases of the forcing. In the following we will present the details of our model and the results of numerical simulations. At the end of this paper we discuss our findings and draw conclusions.

\section{Langevin model including random zonal flows and background turbulence}
Following the work performed by Krommes in Ref. \cite{krommes2000b} on the impact of random flows on the fluctuation levels in simple stochastic models, we consider the passively advected fluctuations of a scalar $\psi$ such as temperature, to obey
\begin{eqnarray}
&&\partial_t \delta\psi(\mathbf{x},t) + \delta \mathbf{V}(\mathbf{x},t). \nabla \delta\psi - D \nabla^2 \delta\psi = \delta f (\mathbf{x},t)
\label{eq1}
\end{eqnarray}

As in Ref. \cite{krommes2000b} to keep the discussion as general as possible the linear physics is modelled by a random external forcing $\delta f$ and a classical dissipation $D \nabla^{2}$. Furthermore, we assume homogeneous statistics and thus only solve eq. (\ref{eq1}) in one dimension, i.e. $ \delta \psi(y,t)$. Here, $\delta \mathbf{V}$ is a statistically specified random flow velocity, corresponding to the random $u(t)$ in eq. (\ref{eq0}). In the present model we ignore the spatial dependence of $\delta \mathbf{V} (\mathbf{x},t)= \delta \mathbf{V}(t)$ employing a Fourier transformation of eq. (\ref{eq1}) we obtain the forced stochastic oscillator equation (see eq. (32) in Ref. \cite{krommes2000b})
\begin{eqnarray}
&&\partial_t \delta\psi_k + i \delta u_k(t) \delta\psi_k + \nu_{k} \delta\psi_k = \delta f_k(t)
\label{Feq1}
\end{eqnarray}
where we assumed the following
\begin{eqnarray}
&& \delta u_k(t) = k_y \bar{V}\exp(i\theta_k(t))\label{eq4}\\
&& \delta f_k(t) = \gamma \exp(i\phi_k(t))\label{eq5}
\end{eqnarray}

In our model, thus, the random flow and forcing are assumed to be similar to oscillators with constant amplitude and phases that varies in time with $\theta_k(t)$ and $\phi_k(t)$, respectively. Here, we used the same definitions for $\gamma\dot= 2\kappa^2 D$, and $\nu_k = k^2 D$ as in Ref. \cite{krommes2000b} with $\kappa$ being the a constant measuring the strength of the forcing, and $k^2=k_y^2$. $\bar{V}$ is a constant measuring the strength of the random flow. Note that this model introduces a multiplicative noise term as well as additive noise term. The scale dependence is introduced by the multiplication by $k_y$ and the prescribed dispersion relation for the natural frequency, see eq. (\ref{eq6}) in the next section. In general one can assume also a stochastic amplitude and therefore treat $\bar{V}$ as a random value with a given statistical property, which would allow for further degrees of freedom in the model. However at this point we aim to study the role of phase self-organisation on the fluctuating scalar and therefore we will assume a constant amplitude.

\section{Phase coupling model}
\begin{figure}
\includegraphics[width=5cm, height=3cm]{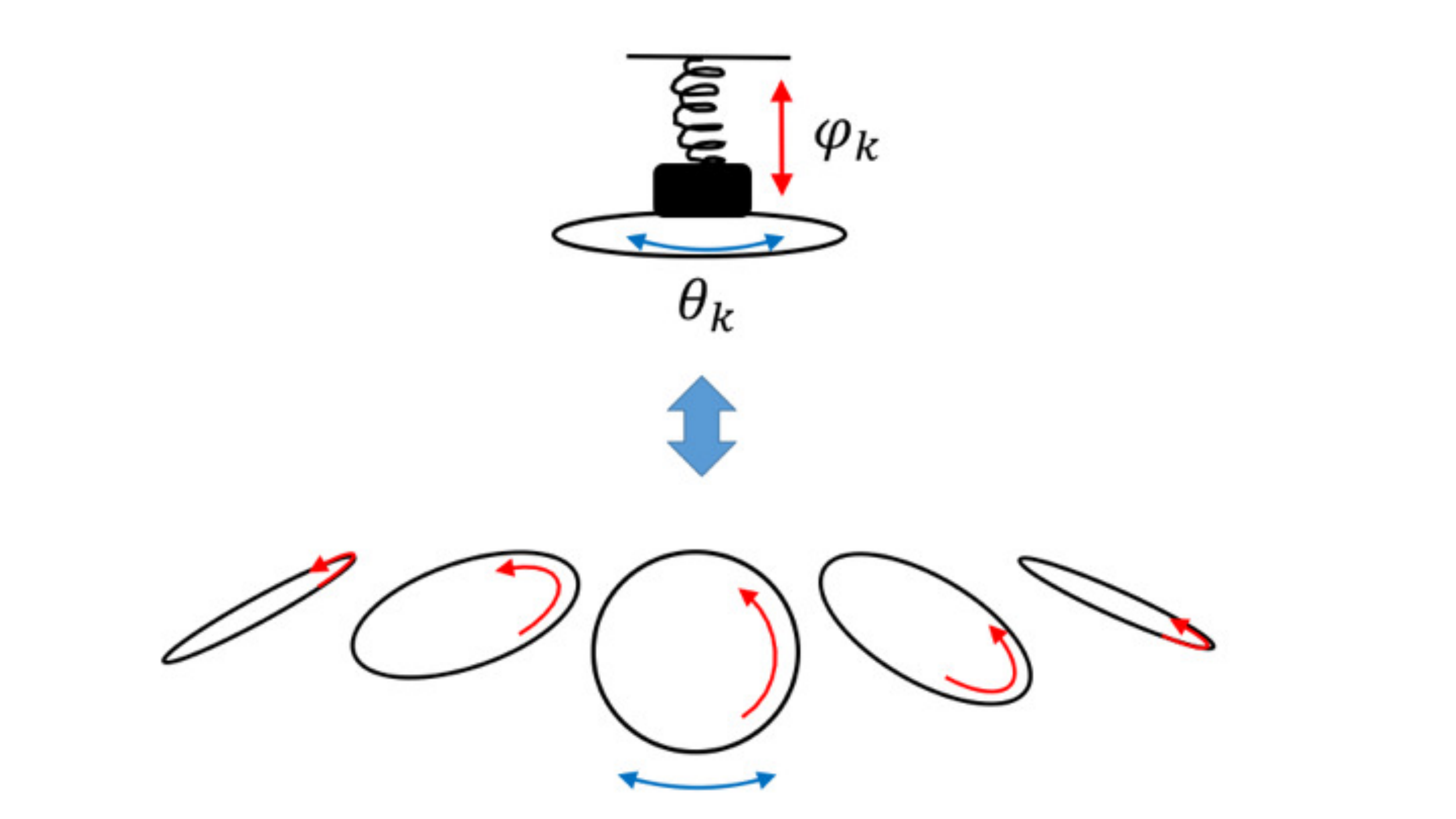}
\caption{\label{fig1} Schematics of the Wilberforce pendulum representing the phase variations corresponding to the oscillating zonal shear modulating the direction of excursion via eddy tilting.}
\end{figure}

We have expanded the original 1D Kuramoto model to a 2D model by assuming that the oscillators perform a two dimensional motion similar to a system of coupled Wilberforce pendulum \cite{wilberforce, moradi2Dmodel} where the motion can be represented by two phases of $\theta$ in longitudinal and $\phi$ in torsional plain, respectively. The dynamics of the phases are described by the two coupled first order differential equations as
\begin{eqnarray}
&&\dot{\theta}_{k}(t)=\omega_k+(2\pi)^{-1}\sum_{i=1}^{N}J_{ik}sin(\theta_i-\theta_k)+\frac{1}{2}\epsilon\phi_{k},
\label{theta1}\\
&&\dot{\phi}_{k}(t)=\zeta_k+(2\pi)^{-1}\sum_{i=1}^{N}S_{ik}sin(\phi_i-\phi_k)-\frac{1}{2}\epsilon\theta_{k},
\label{theta2}\\
&& \;\;\;\;\;\;\;\;\;\; \;\;\;\;\;\;\;\;\;\; \;\;\;\;\;\;\;\;\;\; \;\;\;\;\;\;\;\;\;\; \;\;\;\;\;\;\;\;\;\; \;\;\;\;\;\;\;\;\;\;(k=1,...,N).\nonumber
\end{eqnarray}
where the $\theta_k$ and $\phi_k$ follow a non-linear sinusoidal coupling as in the Kuramoto model \cite{kuramoto} with additional linear cross coupling between the two phases. Here, the analogy is that $\phi$ is the phase corresponding to a fluctuating radial excursion associated with the drift wave, and the $\theta$ is the phase corresponding to the oscillating zonal shear modulating the direction of excursion via eddy tilting \cite{biglaridimondterry1990}, see the representative schematics shown in Fig. \ref{fig1}. The linear cross-coupling introduces a cross correlation between the two motions similar to the regular Lotka-Volterra predator-prey model \cite{lotka1920, volterra1928, goel1971}. This cross-coupling mimics the interactions between the drift wave turbulence and zonal flows which have been observed to follow self-consistent feedback loop systems similar to predator-prey trends as was shown in Refs. \cite{dimond94,Charlton94,dimond2005,dimond2011}. Here, $\dot{\theta}_{k}(t)$, and $\dot{\phi}_{k}(t)$ denote the time derivatives of the phases of $k$th mode. The parameter $\epsilon$ is a free parameter of the model which allows to modify the strength of linear cross-coupling term and $N$ is the number of oscillators or modes considered. $\omega_k$ are the natural frequencies of the flow assumed random and distributed according to a Gaussian distribution with zero mean, $f(\omega)=\exp(-\omega^2/2)/\sqrt 2\pi$, and $\zeta_k$ are the natural frequencies of the forcing, also assumed random and distributed according to a Gaussian distribution, but the mean is prescribed by a spectrum defined through a dispersion relation similar to that of the DWs (see Refs. \cite{Horton,GurcanPRL2009}):
\begin{eqnarray}
&&<\zeta>(k) = -\beta\frac{k}{1+k^2}\label{eq6}
\end{eqnarray}
where $\beta$ is a free parameter of the model. In the usual DW picture this parameter represents a gradient e.g. density gradient, $\delta n/\delta y$.
$J_{ik}$ and $S_{ik}$ measure the strength of the interactions between oscillator $i$ and $k$ in each population, and they are assumed random constants distributed according to a Gaussian distribution with standard deviation defined as $\sigma_{\theta,\phi}=\{F,G\}/(\sqrt(2N))$. Here $F$ and $G$ are control parameters of the model. Note, that in our model low values of $F,\;G$ correspond to weak coupling while high values correspond to strong coupling between the phases in each population.

\section{The numerical set up}
In this work, the numerical integration of eq. (\ref{Feq1}) is performed using the Runge-Kutta 4th order scheme (RK4) with time stepping length $\delta t=2\pi\times dt$ where $dt$ is the optimum time interval varying for each integration while the sampling time step is $\Delta t=0.01$. At each time step the values of $ \delta \omega_k(t)$ and $\delta f_k(t)$ are updated through numerical integration of eqs. (\ref{theta1} and \ref{theta2}) using RK4. For initial conditions we use $|\delta\psi_k(0)|=1$, with the phases of $\delta\psi_k(0)$ set to zero. For the random flow and the forcing we set the initial phases as $\theta_k(0)=\phi_k(0)=0$. The mode number $k$ is chosen following a shell model type approach by setting $k_n = k_0 \times g^{n}$ where $n=1,\dots N/5$ where $N=125$ corresponds to the number of modes, with $k_0=1$ and $g=1.25$. Each mode number $k$ is represented by 5 oscillators with different natural frequencies distributed according to a Gaussian distribution around a mean value prescribed by the dispersion relation given by eq. (\ref{eq6}) and we employ an averaging over these 5 modes. An averaging over $N_{s}=10$ samples of $J_{ik}, S_{ik}$ is also performed. In the present study, the time span of simulations is of the order of $t = 20$ which allows for the system to relax to a steady state. 

An analytic expression for the order parameter $Z(t)=\sum_{k=1}^{N} \exp(i\theta_{k})/N$ was derived by Kuramoto that describes the quality of the synchronisation of the ensemble of oscillators with $0\le Z\le1$. Here, $Z=0$ corresponds to a complete a-synchronised state while $Z=1$ corresponds to a total synchronised state. We have calculated the values of the order parameter separately for each $\theta$ and $\phi$ phases, and averaged over $N_s$ samples denoted by $[Z_{\theta,\phi}(t)]$. 


\section{Results of numerical simulations} 
In the following we present the results of the numerical integration of eqs. (\ref{Feq1}-\ref{theta2}) for various phase coupling parameters. To measure the dynamic of the system we computed the evolution of the energy like quantity 
\begin{eqnarray}
&&C(t,t') = <[|\delta \psi_k(t)\delta\psi^{*}_k(t')|]>
\label{corr}
\end{eqnarray}
where $[...]$ represents the averaging with respect to $k$ and $<...>$ denotes the sample averaging over a number of different realisations of random values: $\omega_k$, $\zeta_k$, $J_{ik}$ and $S_{ik}$. 

\subsection{The case without the linear cross-coupling} 
At first we examine the role of phase synchronisation in the absence of the linear cross-coupling between the flow and the forcing by setting $\epsilon=0$, see eqs. (\ref{theta1} and \ref{theta2}). 

\begin{figure}
\includegraphics[width=4cm, height=3cm]{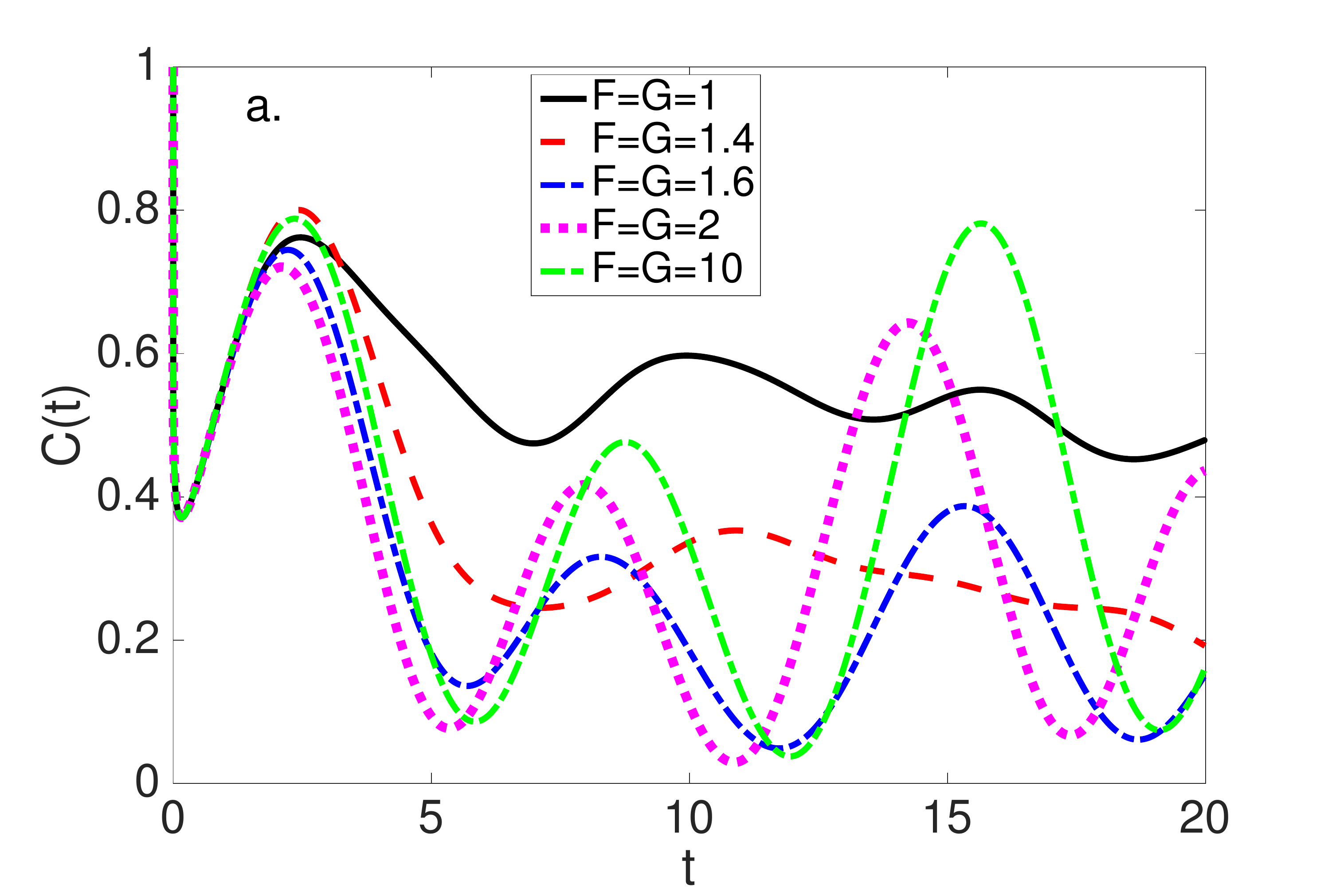}\includegraphics[width=4cm, height=3cm]{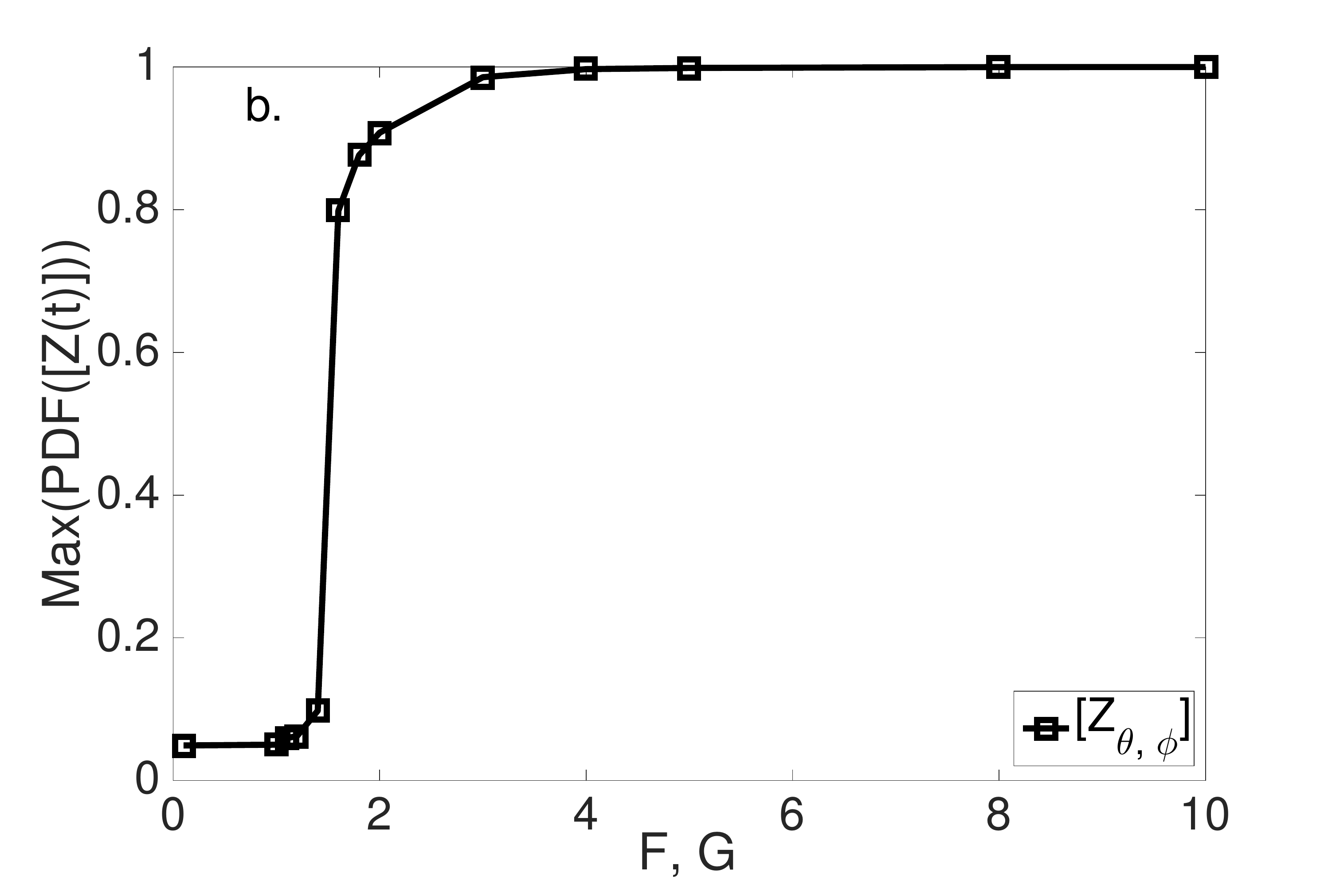}
\caption{\label{fig2} (a) Time evolution of sample averaged single time auto-correlation $C(t)$ for different control parameters ($F$ and $G$). (b) The maximum values of the PDFs of sample averaged order parameter $[|Z_{\theta, \phi}|]$ as function of control parameters. The parameters of the numerical simulation are $\beta=0$, $\epsilon=0$, $N=125$, $N_s=10$, $\Delta t=0.01$, $\gamma=1$, $\bar{V}=0.01$, $\nu=0.01$, $g=1.25$, and $k_0=1$.}
\end{figure}

\begin{figure}
\includegraphics[width=4cm, height=6cm]{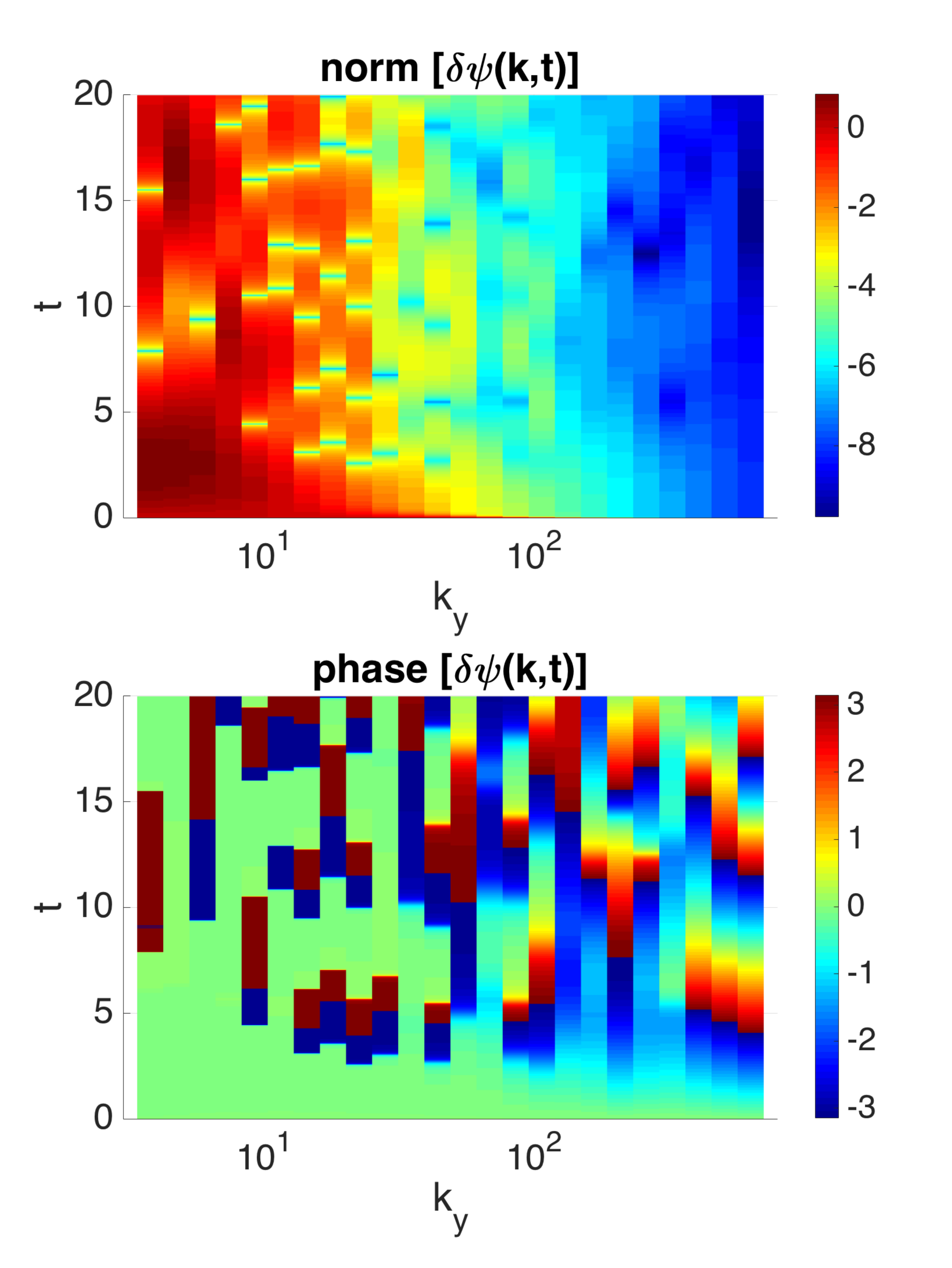}\includegraphics[width=4cm, height=6cm]{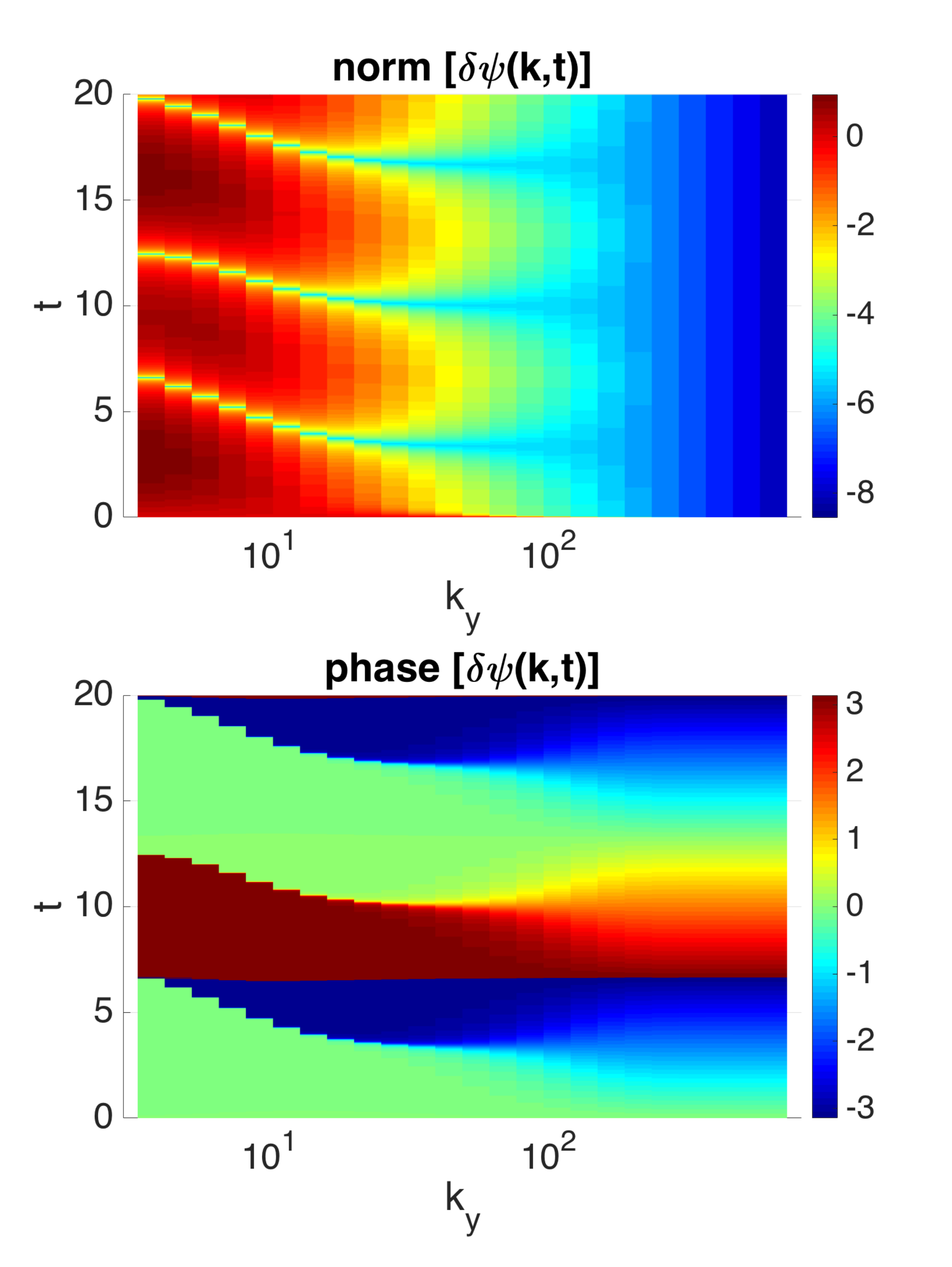}
\caption{\label{fig3} (top) The logarithm of norm and (bottom) the phase of $\delta\psi_k$ as functions of mode number $k$, and time $t$. Left figures present the results for the $F=G=1$ and right figures present the computed values for $F=G=10$. }
\end{figure}

Figure \ref{fig2} (a) shows the impact of synchronisation of the phases in $\theta_k$ and $\phi_k$ populations on the time evolution of the auto-correlation function $C(t)$. Here $\beta=0$, and the synchronisation states are controlled by the parameters $F$ and $G$, as seen in Fig. \ref{fig2} (b) where the maximum of the PDF of the sample averaged $[Z(t)]$ are shown. In agreement with the well-known Kuramoto model we find that by increasing the control parameters the phases in each population move from an a-synchronised ($[Z(t)]\sim0$) to a synchronised state ($[Z(t)]\sim1$). This shift from a-synchrony to synchrony significantly enhances the evolution of the $C(t)$. For an a-synchronised condition after an initial decay due to dispersion term, $C(t)$ saturates to a finite level around $C(t) \sim 0.5$. As the phases become more and more synchronised $C(t)$ saturates to an oscillatory state with two amplitudes of $\Delta C(t)\sim 0.4$ and $\sim 0.7$. The dynamics of the norm and the phase of $\delta \psi_k$ for the two extreme conditions, i.e. $F=G=1$, and $F=G=10$ are presented in Figs. \ref{fig3} left and right, respectively. A clear distinction between the two cases is observed where in the synchronised system, coherent oscillations appear across the $k$ space in both the norm and the phases of $\delta \psi_k$, see Fig. \ref{fig3} right. Here, we find that for the low $k$, phases oscillate between $0$ and $\pi$ with a low frequency, while the phases of the higher $k$ modes try to catch on. The norm also shows oscillations with similar frequency and the minima correspond to when the phases reach the value $0$.

\begin{figure}
\includegraphics[width=4cm, height=3cm]{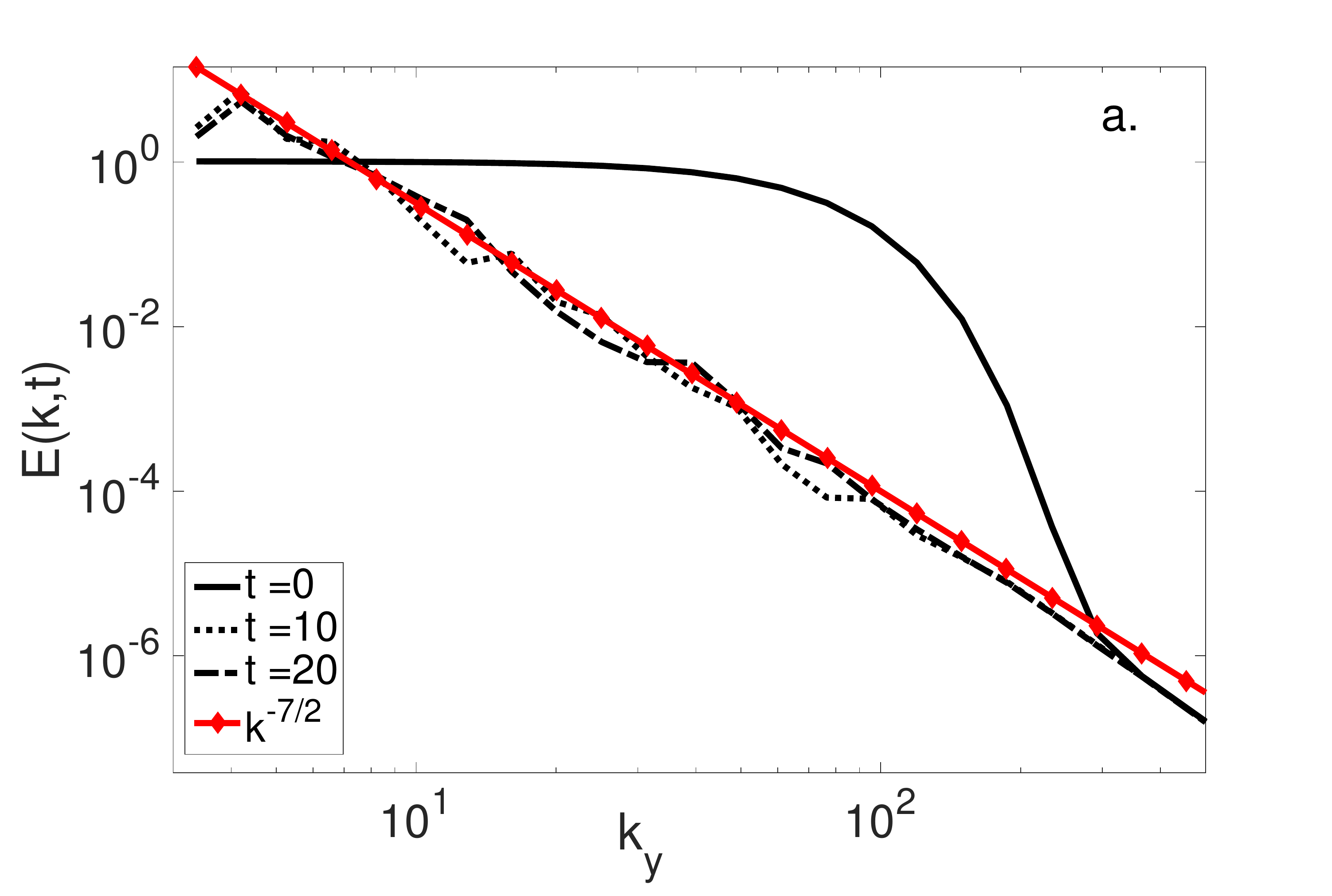}\includegraphics[width=4cm, height=3cm]{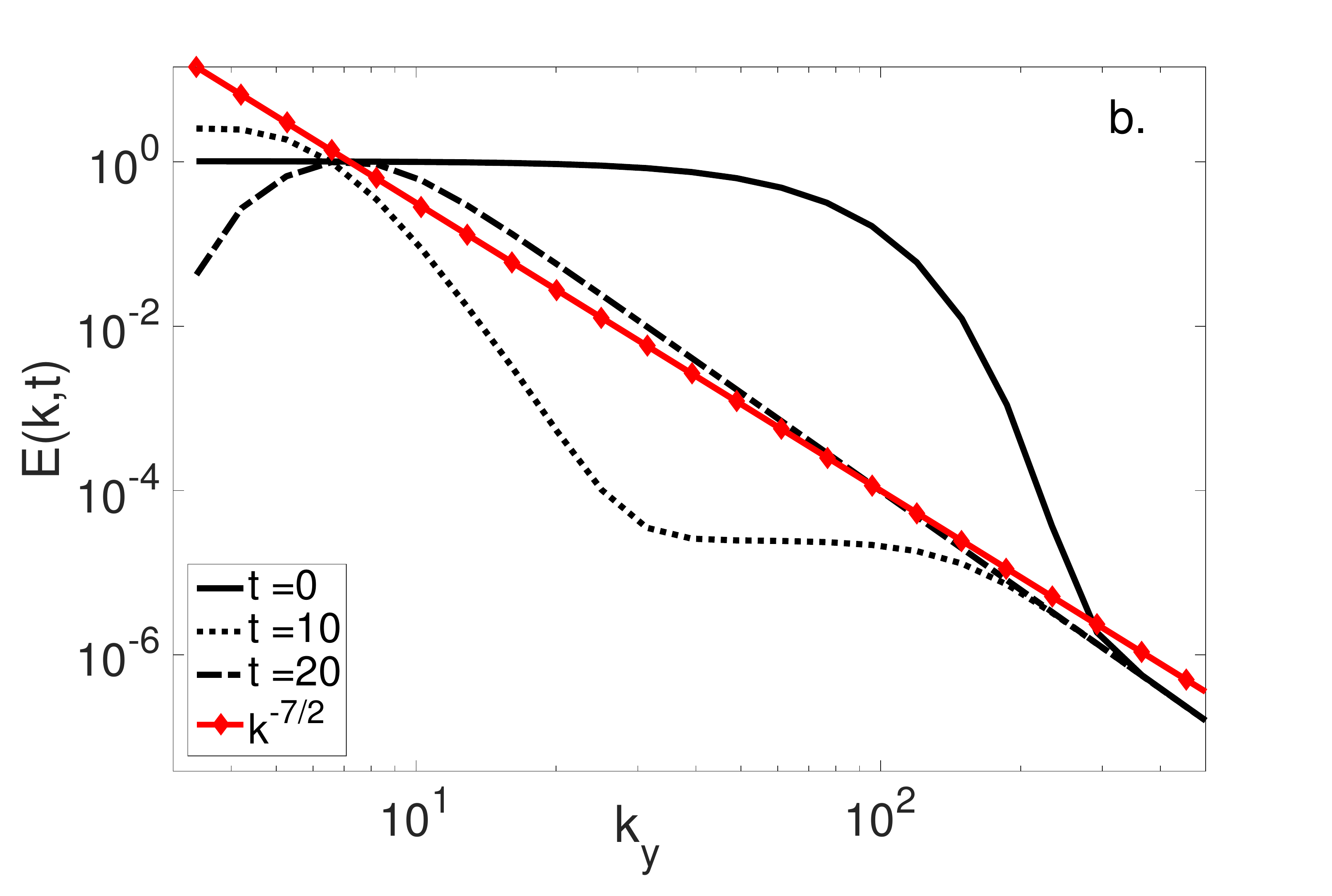}
\caption{\label{fig4} Sample averaged energy spectrum $E(k)$ as functions of mode number $k$ and at three time slices of $t=0$ (solid lines), $t=10$ (dotted lines) and $t=20$ (dashed dotted lines). (a) For the a-synchronised ($F=G=10$) and (b) the synchronised ($F=G=1$) $\theta$ and $\phi$ populations. The red lines with symbols present $k^{-7/2}$ power law scaling.}
\end{figure}

\begin{figure}
\includegraphics[width=4cm, height=3cm]{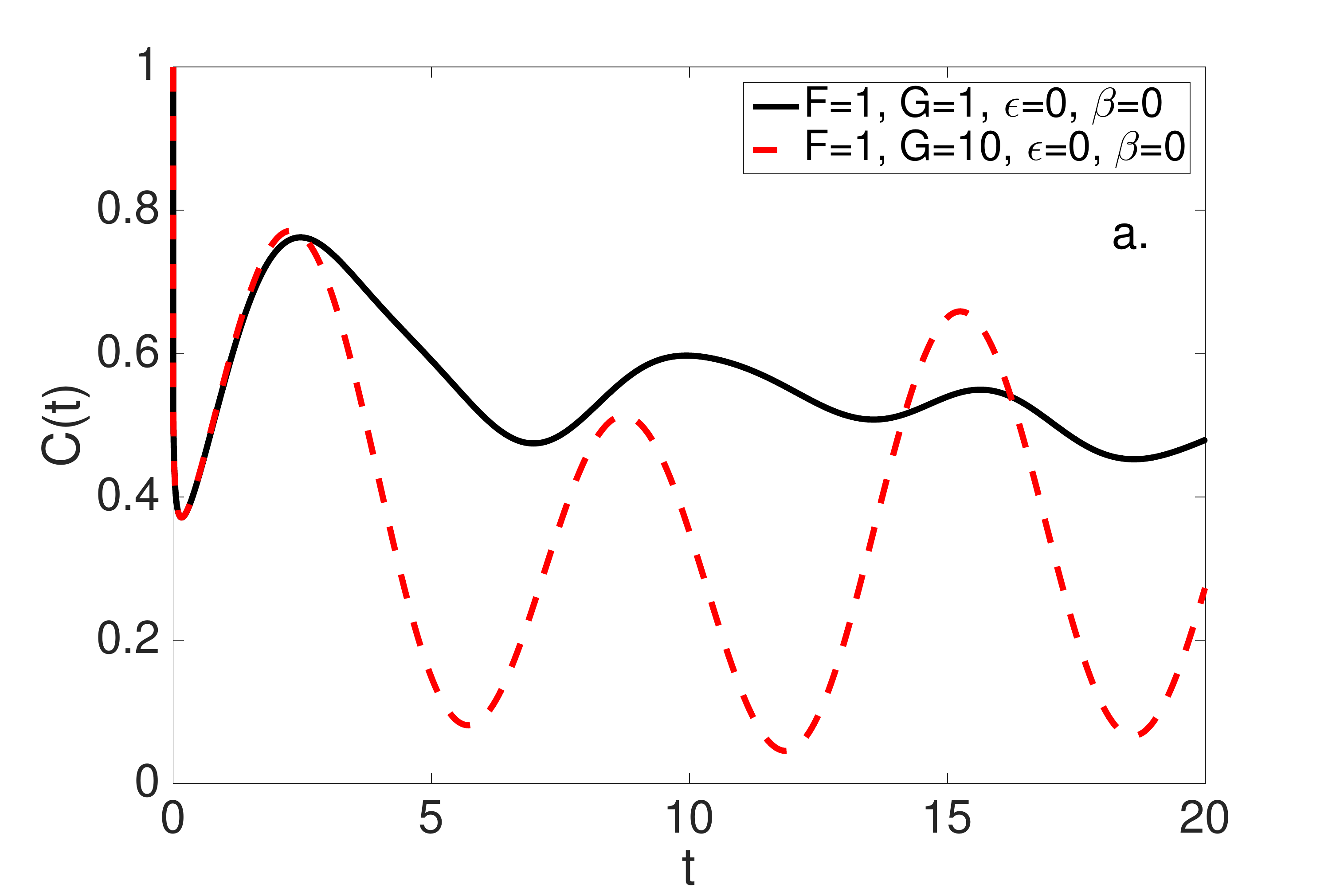}\includegraphics[width=4cm, height=3cm]{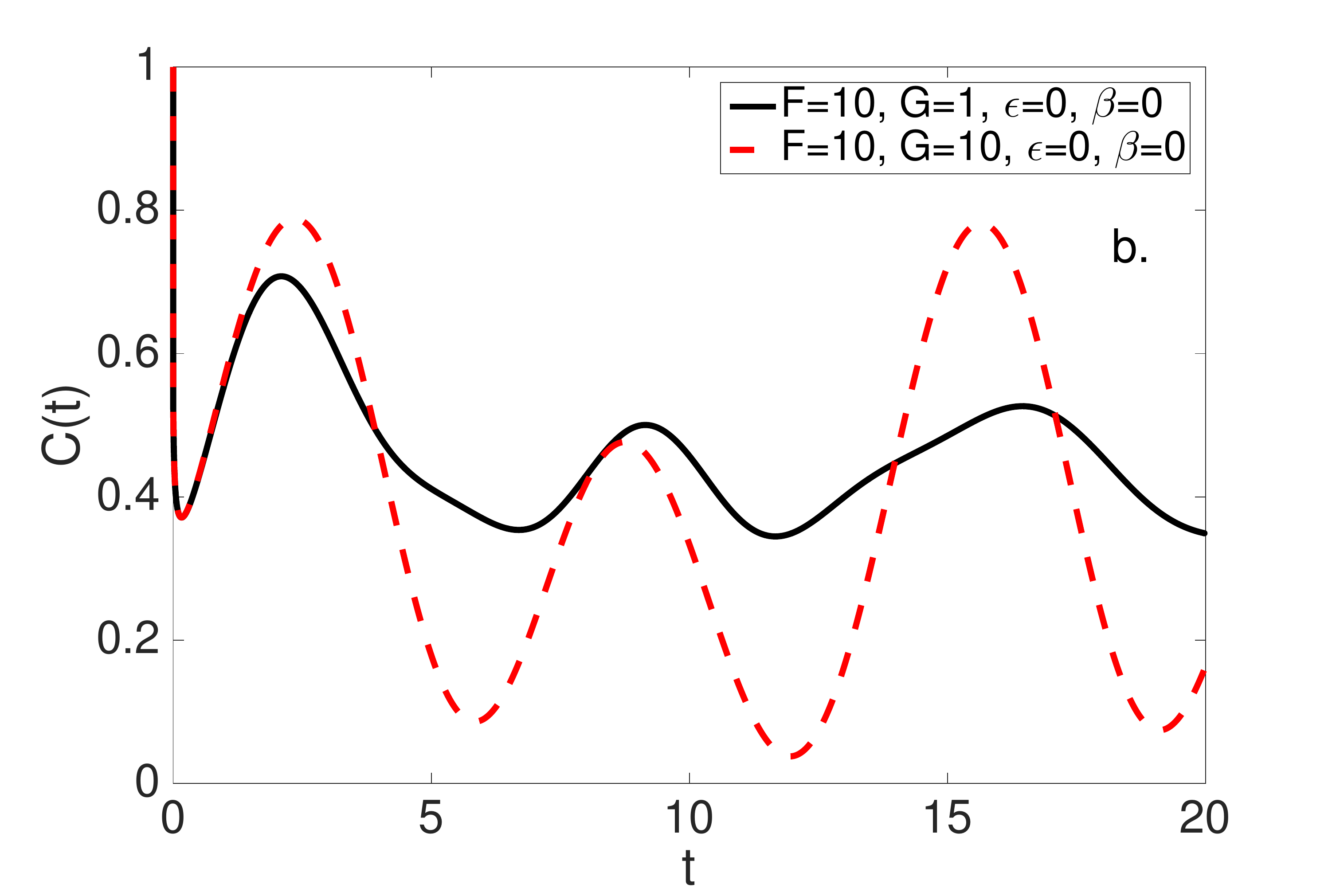}
\caption{\label{fig5} Time evolution of sample averaged $C(t)$ for different control parameters (a) $F=G=1$ (solid line) and $F=1, G=10$ (dashed line). (b) $F=10, G=1$ (solid line) and $F=G=10$ (dashed line).}
\end{figure}

The averaged energy spectrum, $E(k,t)=1/2\sum_{k'=1}^{5}|\delta\psi_{k'}(t)|^2/5$, at three different times are shown in Figs. \ref{fig4} (a) for the a-synchronised, and (b) the synchronised cases of Figs. \ref{fig3}. The a-synchronised spectrum follows a power law decay of the form $k^{-7/2}$ while the synchronised spectrum has a faster decay rate for the low $k$ and a plateau region for the mid-range in $k$ at the time when the phases of $\delta \psi_k$ are at $0$. The dynamic of the $\delta\psi_k$ is strongly affected by the phase states of the forcing, while little dependence between synchronised and a-synchronised phase states of the flow are observed, see Figs. \ref{fig5} (a and b).

\begin{figure}
\includegraphics[width=4cm, height=3cm]{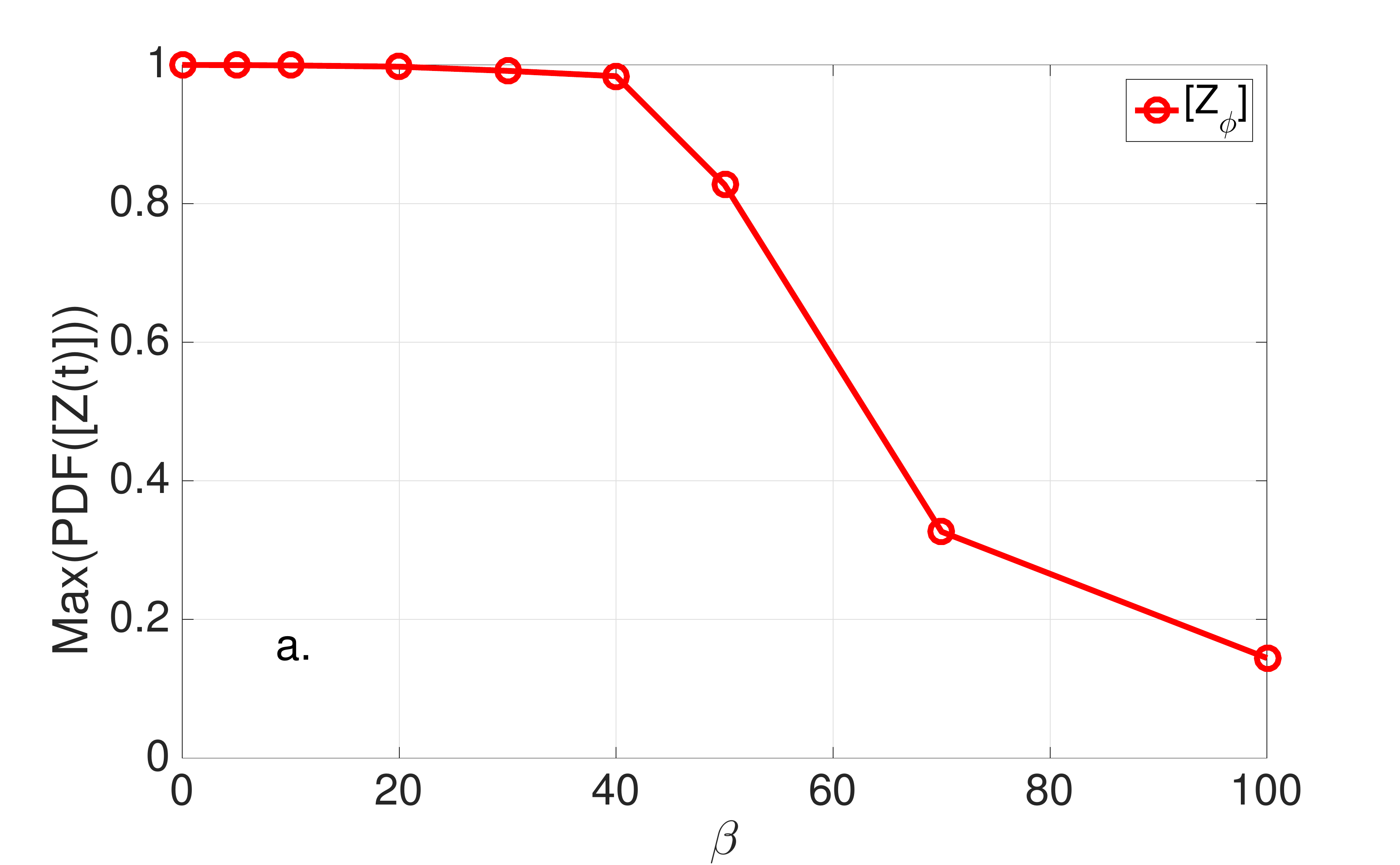}\includegraphics[width=4cm, height=3cm]{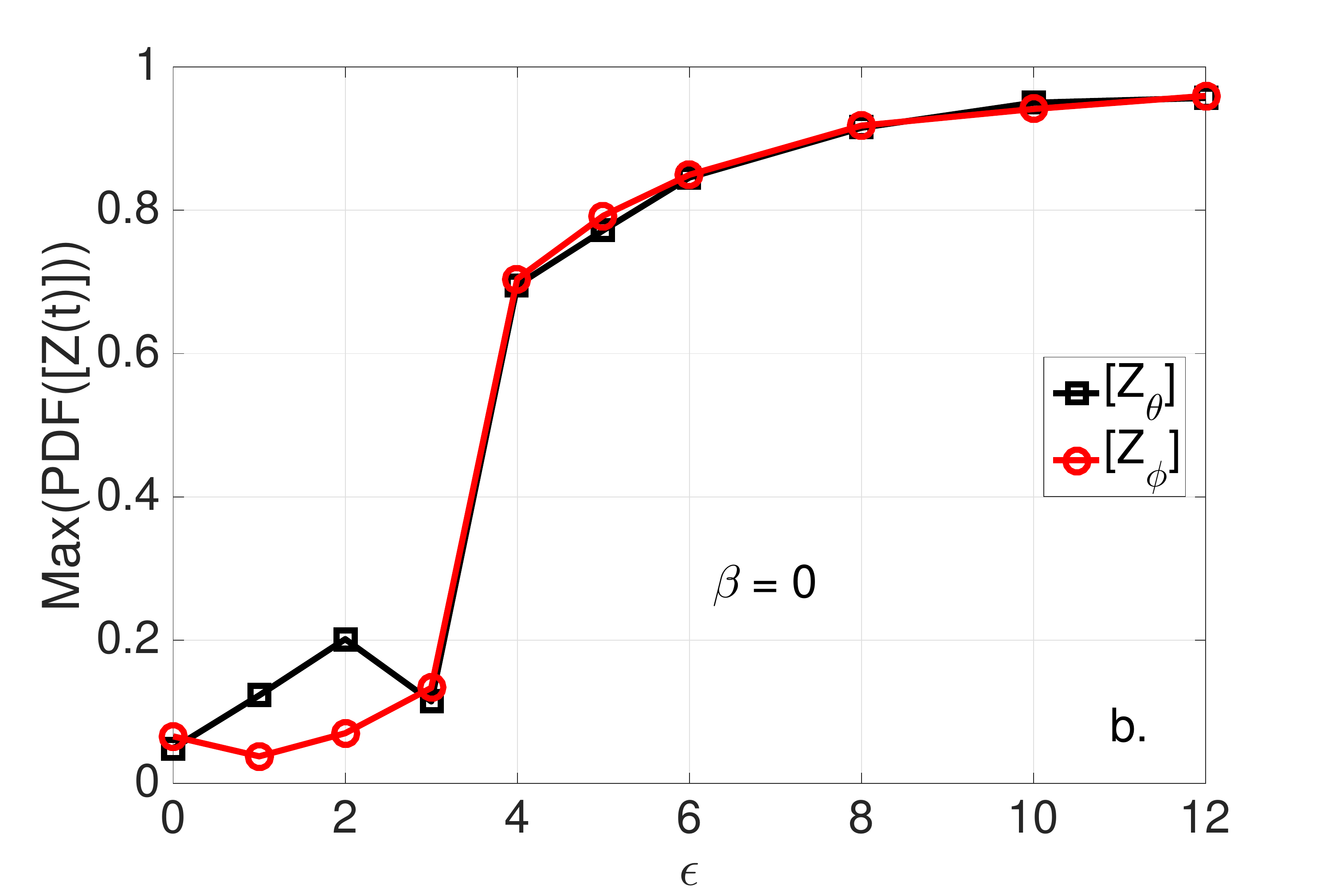}\\
\includegraphics[width=4cm, height=3cm]{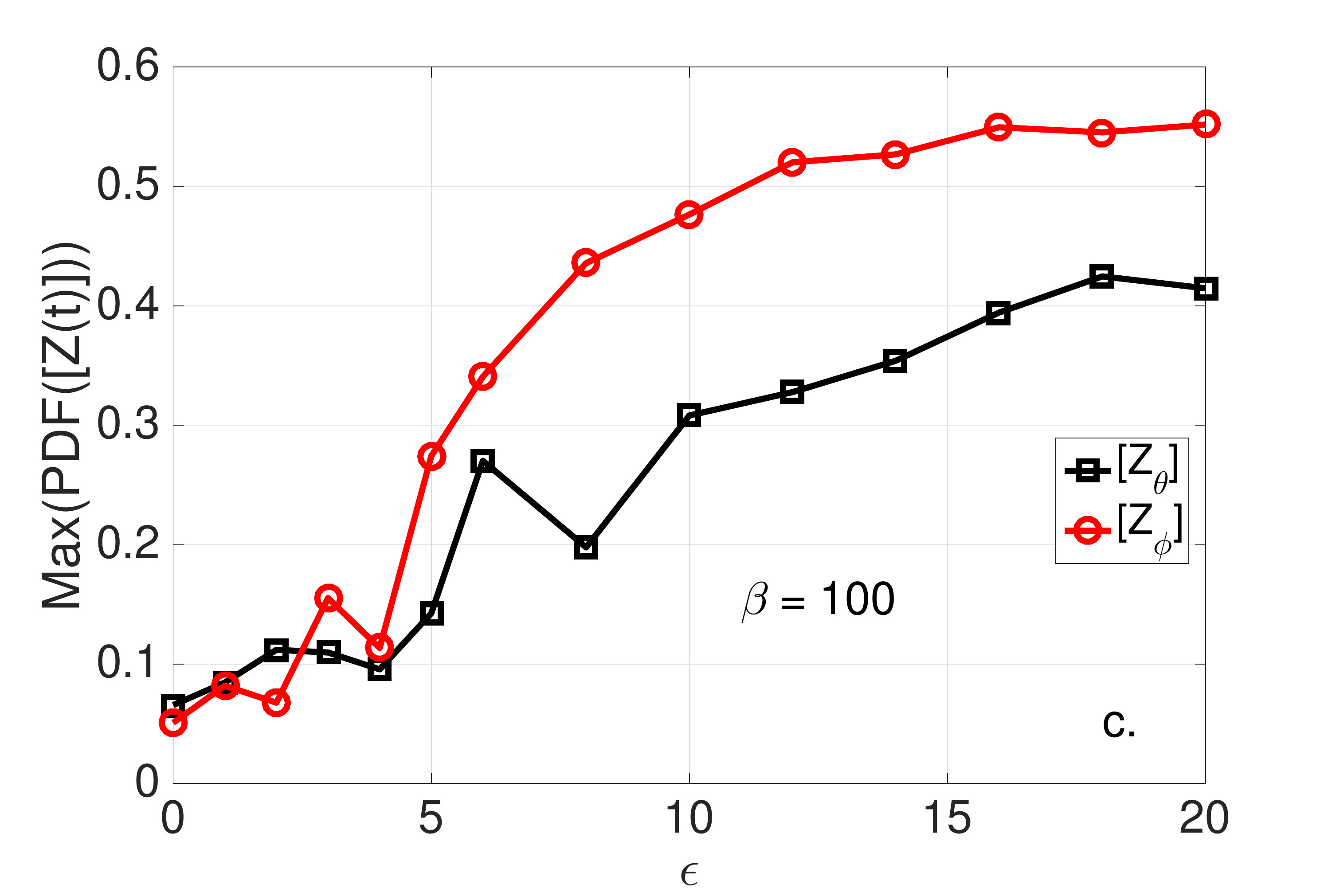}\includegraphics[width=4cm, height=3cm]{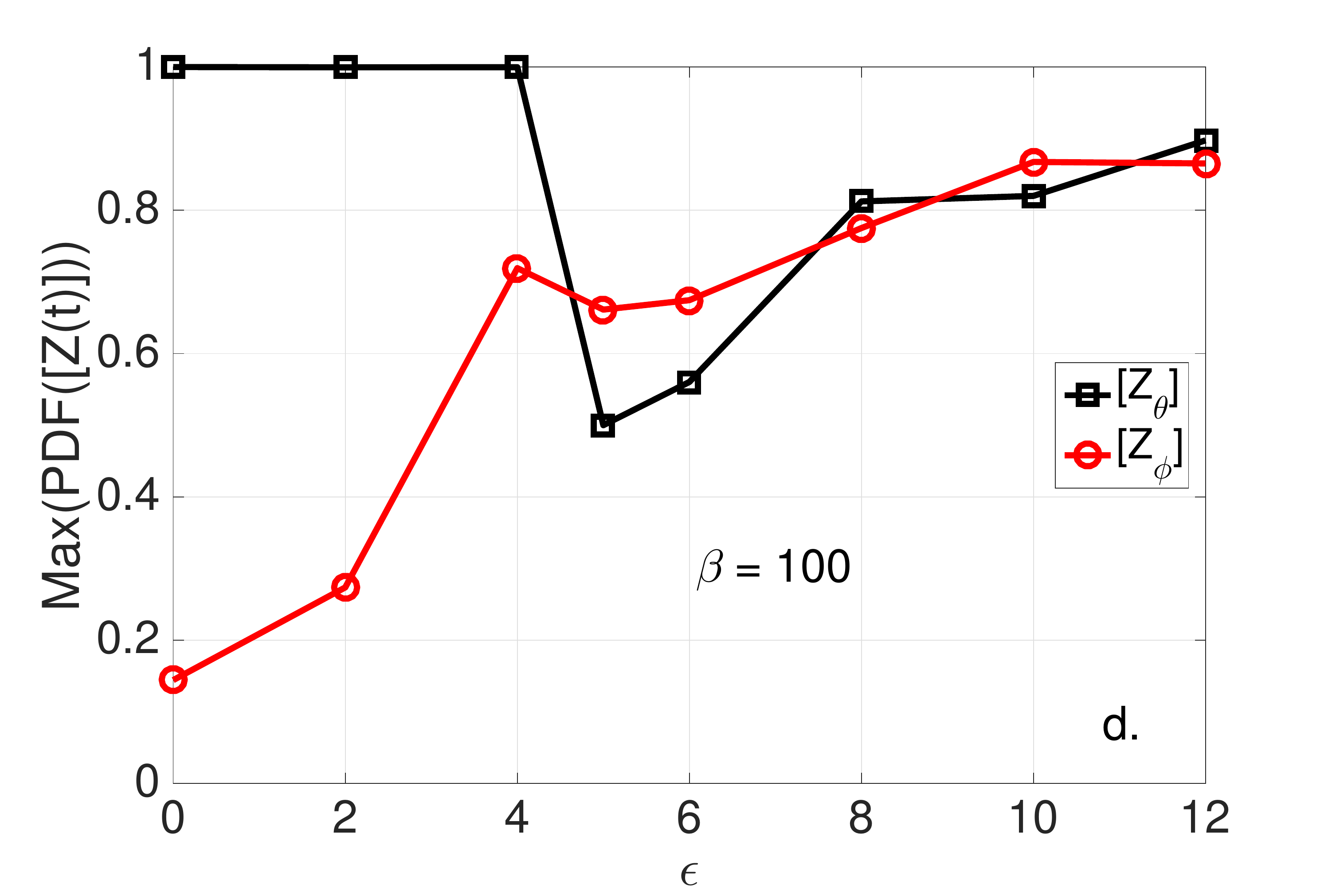}
\caption{\label{fig6} (a) The maximum of the $[Z(t)]$ PDFs as function of $\beta$ for the case without linear cross-coupling ($\epsilon=0$). The maximum of the $[Z(t)]$ PDFs as functions of $\epsilon$ with $\beta=0$ where $F=G=1$ (b), and with $\beta=100$ for (c) $F=G=1$, and (d) $F=G=10$.}
\end{figure} 

\begin{figure}
\includegraphics[width=4cm, height=3cm]{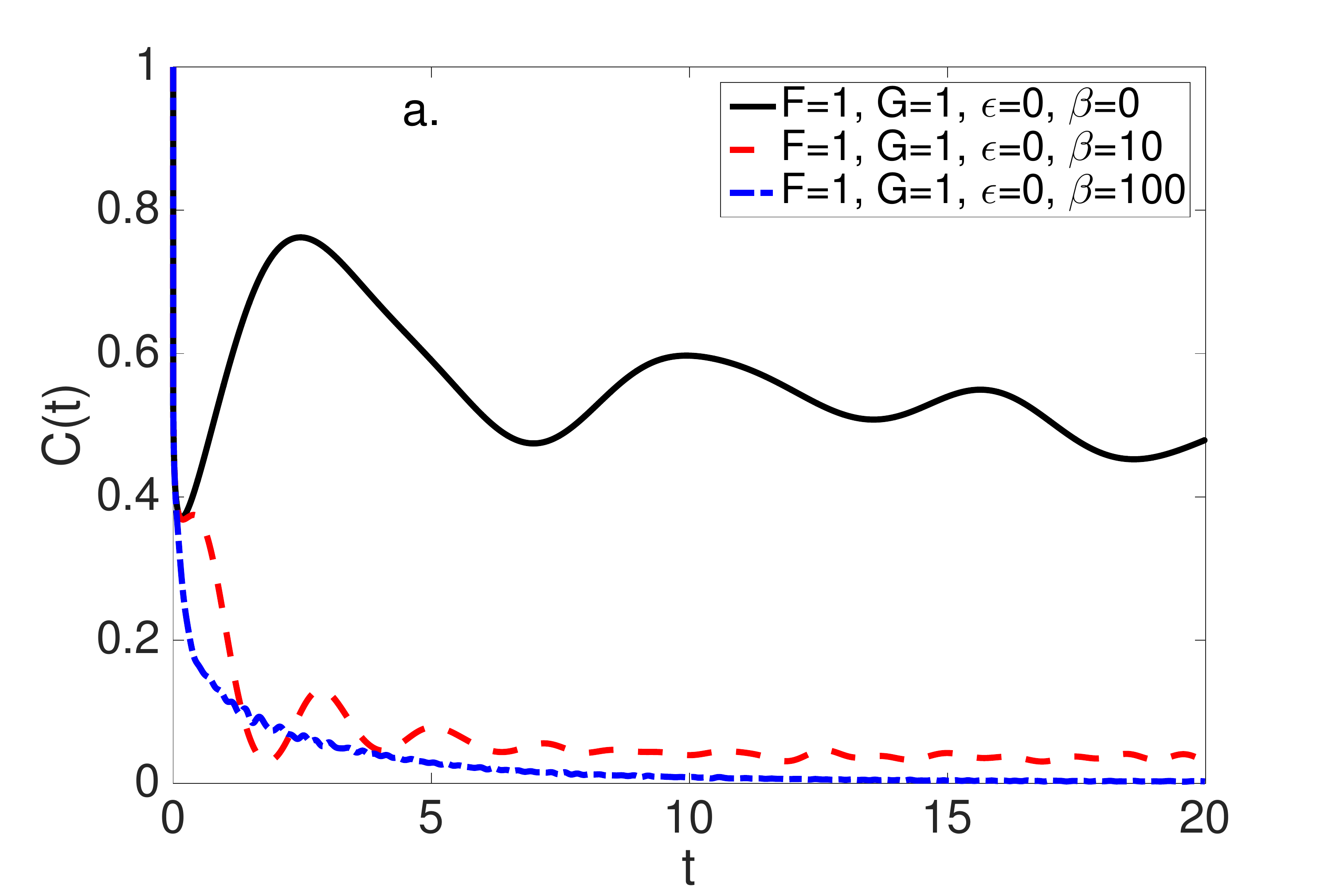}\includegraphics[width=4cm, height=3cm]{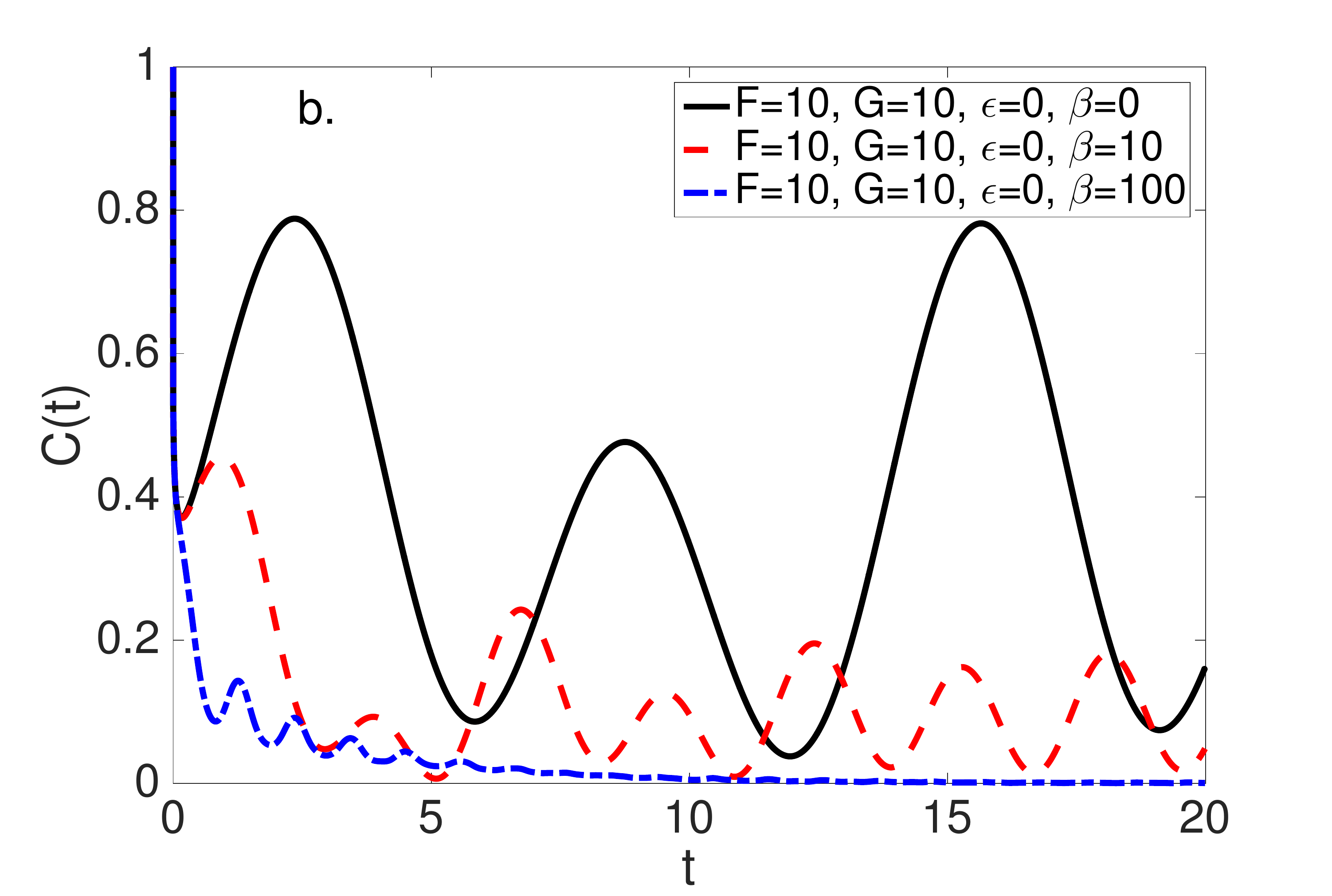}
\caption{\label{fig7} Time evolution of sample averaged $C(t)$ for different $\beta$ values for (a) $F=G=1$ (b) $F=1, G=10$. (Solid lines) show the results for $\beta=0$, (dashed lines) $\beta=10$ and (dashed-dotted lines) $\beta=100$.}
\end{figure}

\begin{figure}
\includegraphics[width=4cm, height=6cm]{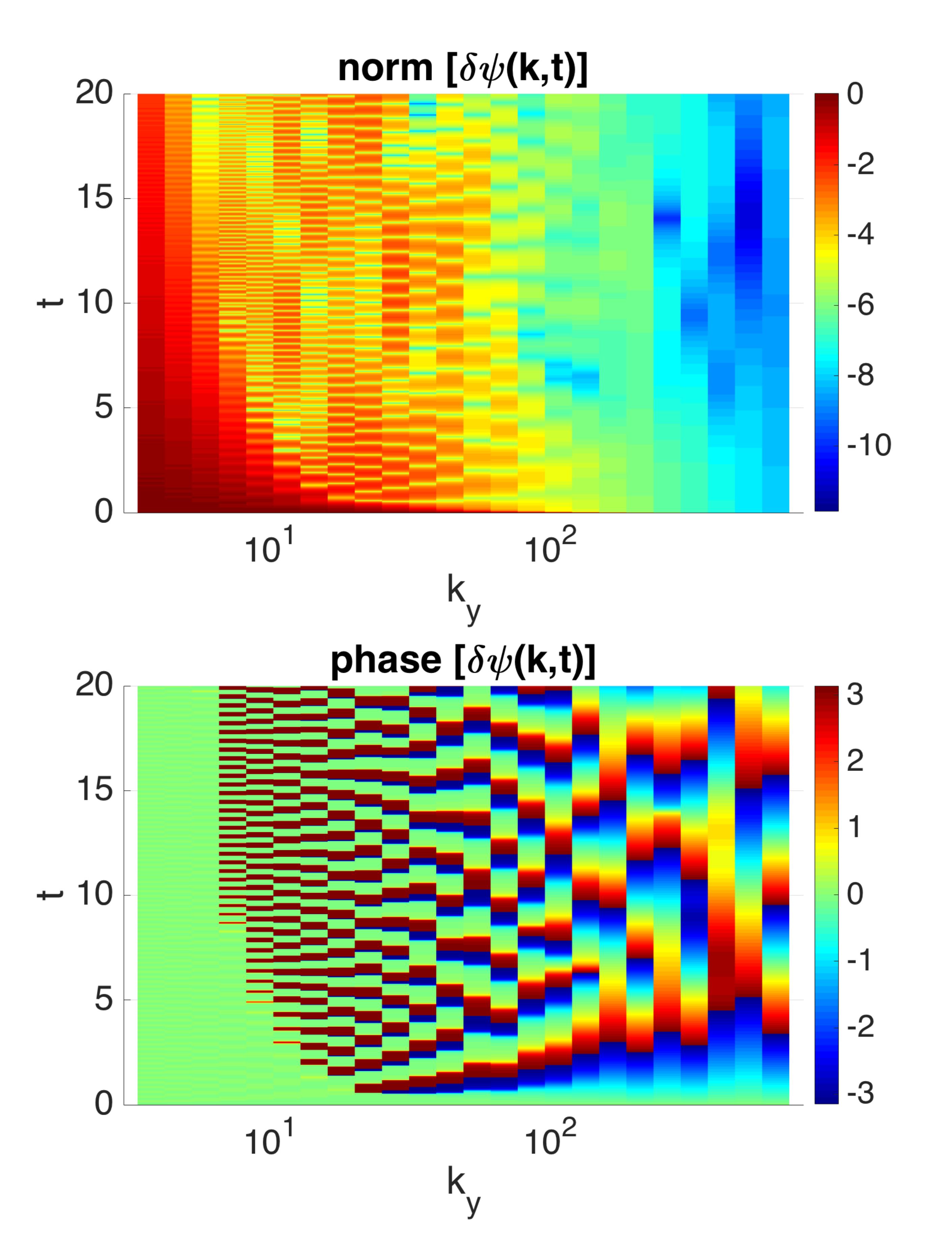}\includegraphics[width=4cm, height=6cm]{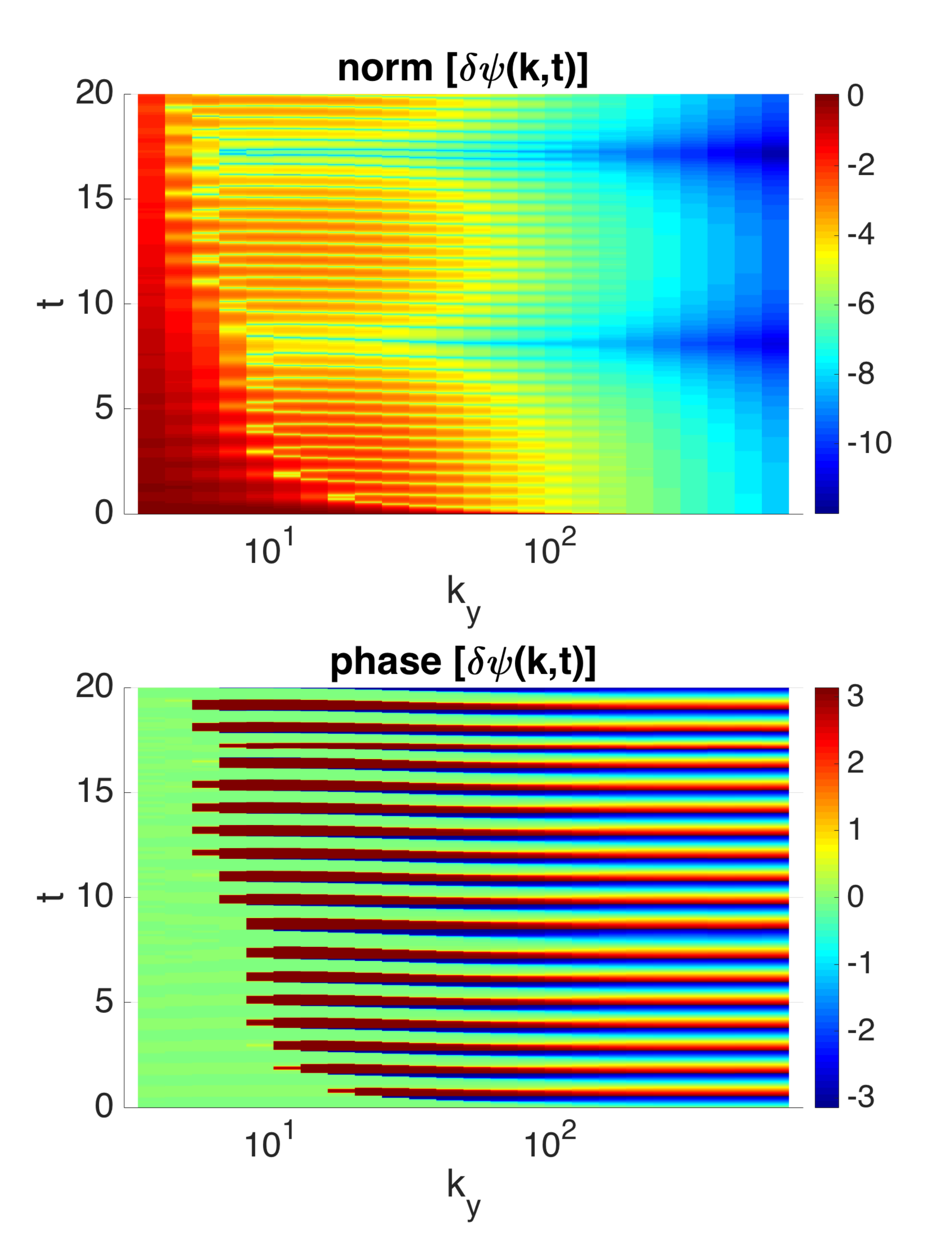}
\caption{\label{fig8} (top) The logarithm of norm and (bottom) the phase of $\delta\psi_k$ as functions of mode number $k$, and time $t$. Left figures present the results for the $F=G=1$ and right figures present the computed values for $F=G=10$. Here, $\beta=100$ and $\epsilon=0$.}
\end{figure}

\begin{figure}
\includegraphics[width=4cm, height=3cm]{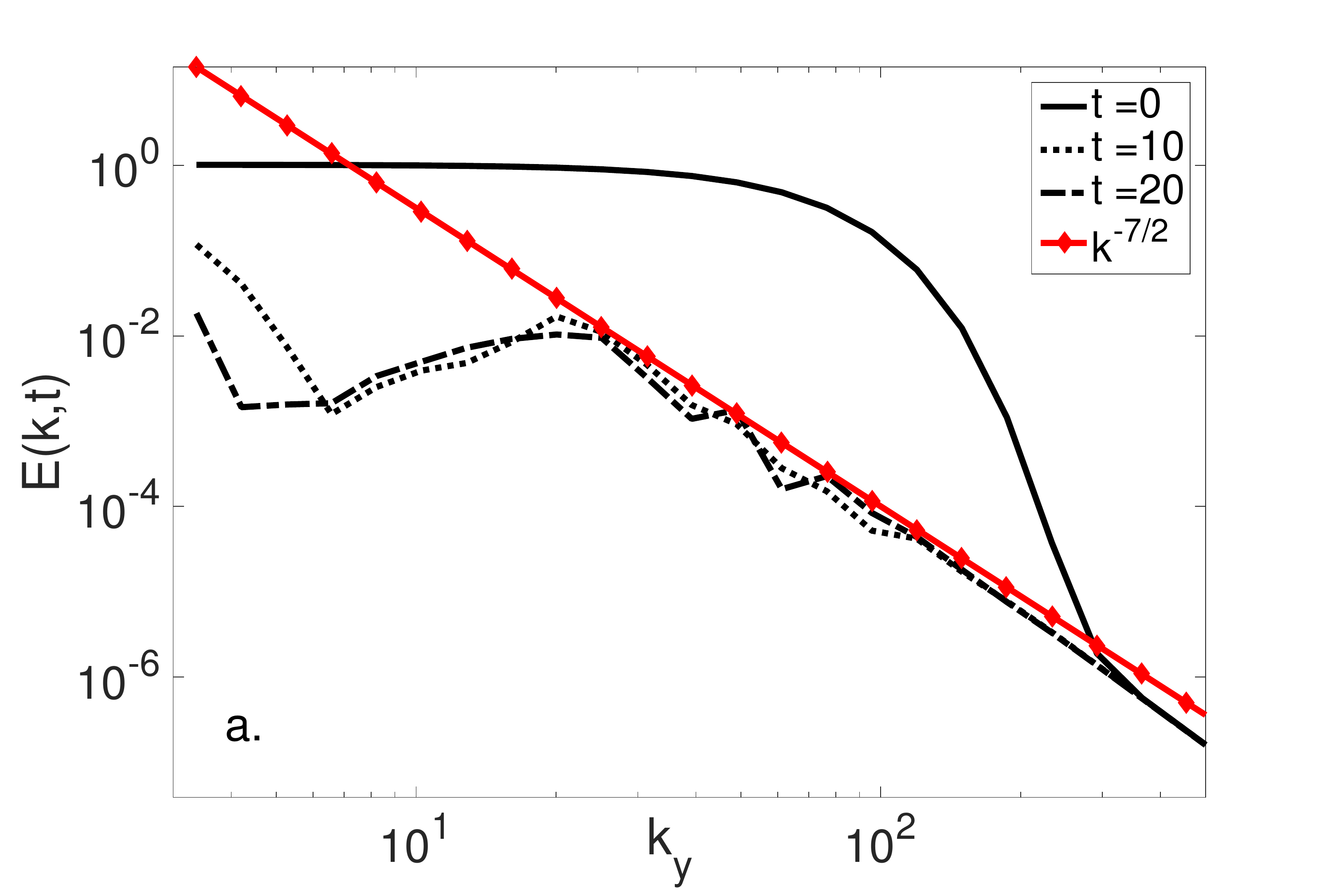}\includegraphics[width=4cm, height=3cm]{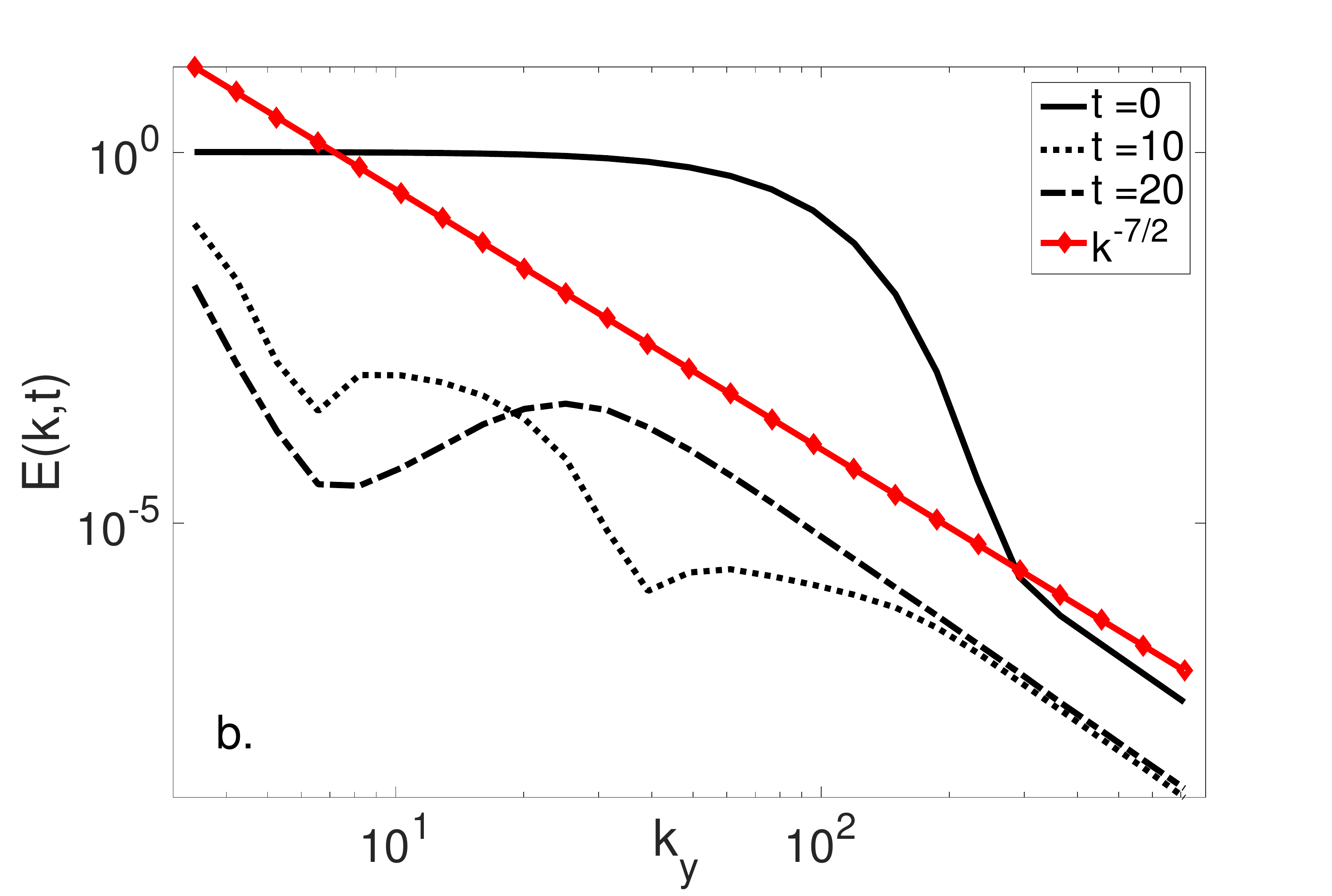}
\caption{\label{fig9} Sample averaged energy spectrum $E(k)$ as functions of mode number $k$ and at three time slices of $t=0$ (solid lines), $t=10$ (dotted lines) and $t=20$ (dashed dotted lines). (a) For the a-synchronised ($F=G=10$) and (b) the synchronised ($F=G=1$) $\theta$ and $\phi$ populations. The red lines with symbols present $k^{-7/2}$ power law scaling. Here, $\beta=100$ and $\epsilon=0$.}
\end{figure}

Figures \ref{fig6} (a-d) show the sensitivity scans of the phase eqs. \ref{theta1} and \ref{theta2} on the different free parameters of model. From Fig. \ref{fig6} (a and b) it is clear that by introducing a dispersion relation of the form given in eq. (\ref{eq6}) the phases of the forcing desynchronise. The linear cross-coupling of the two populations, however, results in synchronisation even for control parameters of an a-synchronised state, i.e. $F=G=1$. For $\epsilon>5$ the synchronisation effect of the cross-coupling term can gain over the desynchonization effect of the dispersion relation, as seen in Figs. \ref{fig6} (c and d).  

The impact of a scale dependence for the natural frequencies via a dispersion relation on the dynamic of $\delta\psi_k$ is two fold. On the one hand an increase in $\beta$ results in a strong reduction of the fluctuation amplitude auto-correlation, $C(t)$. On the other hand as was shown in Fig. \ref{fig6} (a) with increasing $\beta$ the phases of the forcing desynchronise and therefore, the solutions of the cases with $F=G=1$ and $F=G=10$ converge towards the same saturated level, see Figs. \ref{fig7} (a and b). Figures \ref{fig8} illustrate the computed time evolutions of the norm and the phases of $\delta\psi_k$ for $F=G=1$ (left figures) and $F=G=10$ (right figures) with $\beta =100$. Comparing to the $\beta=0$ results shown in Figs. \ref{fig3} however, an increase in $\beta$ seems to synchronise the phases of $\delta \psi_k$, and increase their frequency in both synchronised and a-synchronised phase states of the flow and the forcing. The norm of $\delta \psi_k$ on the other hand shows strong decay in time specially for low $k$ modes, with the oscillations breaking up for the synchronised case, see Fig. \ref{fig8} top right.

The effect of increasing $\beta$ on the time averaged energy spectrum, $E(k,t)$ at three time slices, are illustrated in Figs. \ref{fig9} (a and b). From this figure it is clear that by increasing $\beta$ the energy at lower $k$ is decreased, and the spectrum moves away from $E\propto k^{-7/3}$ scaling. In the case with a-synchronised $\theta_k,~\phi_k$ ($F=G=10$), however, the energy spectrum obeys a non-monotonic change as function of $k$ with its peak located at medium $k$ range.      

In summary, our results show that if one takes into account phase dynamics for the flow and the forcing the assumption of a fully stochastic phase state for the $\delta \psi_k$ is more relevant for high $\beta$ values while for lower $\beta$ the a-synchronised and synchronised phases differ significantly, and one could expect the formation of coherent modulations in the latter case. 

\begin{figure}
\includegraphics[width=4cm, height=3cm]{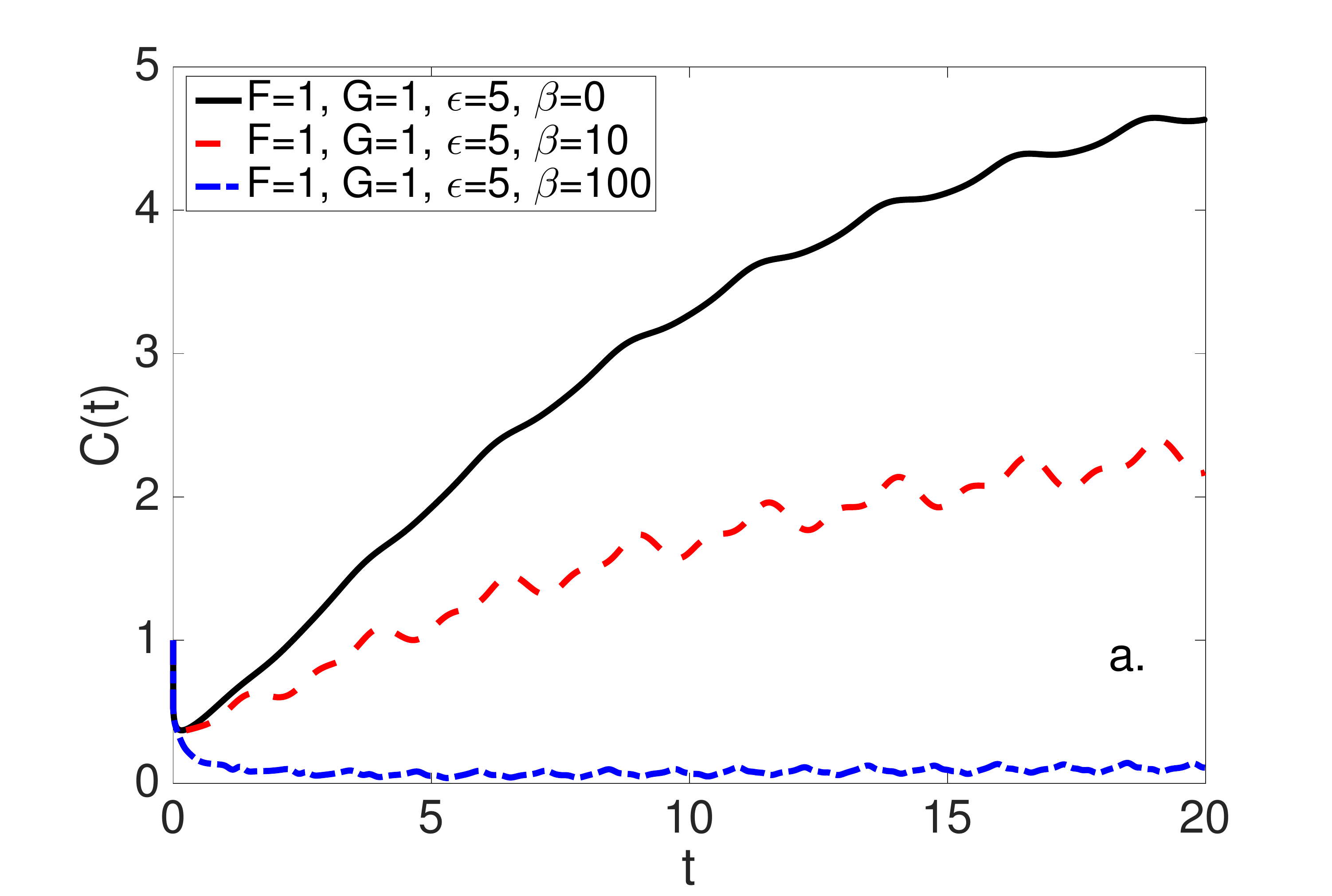}\includegraphics[width=4cm, height=3cm]{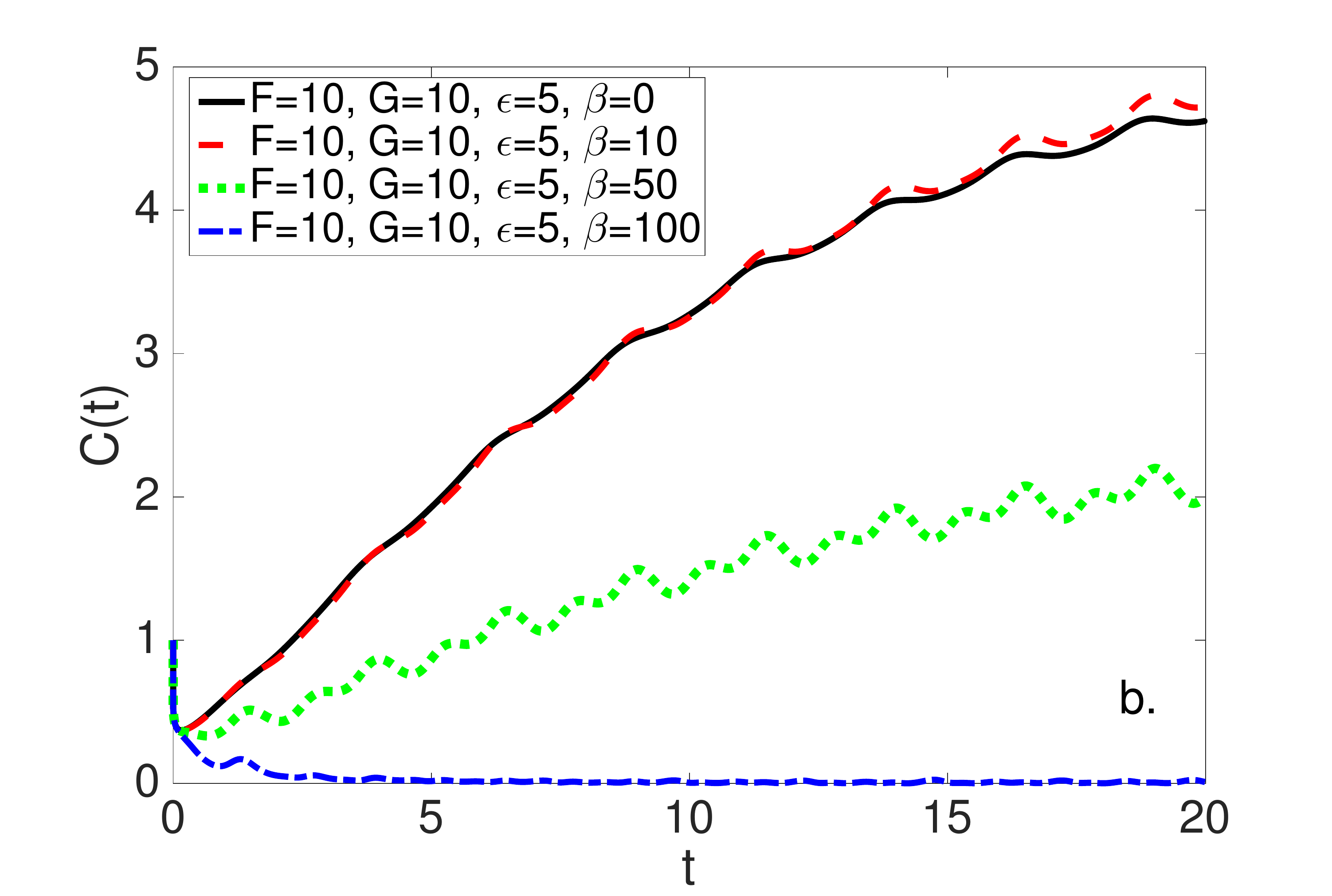}
\caption{\label{fig10} (a) Time evolution of sample averaged $C(t)$ for different $\beta$ at $\epsilon=5$ and (a) $F=G=1$ (b) $F=1, G=10$. (Solid lines) show the results for $\beta=0$, (dashed lines) $\beta=10$, and (dashed-dotted lines) $\beta=100$. In (b) (green dotted line) shows the case with $\beta= 50$.}
\end{figure}

\begin{figure}
\includegraphics[width=6cm, height=8cm]{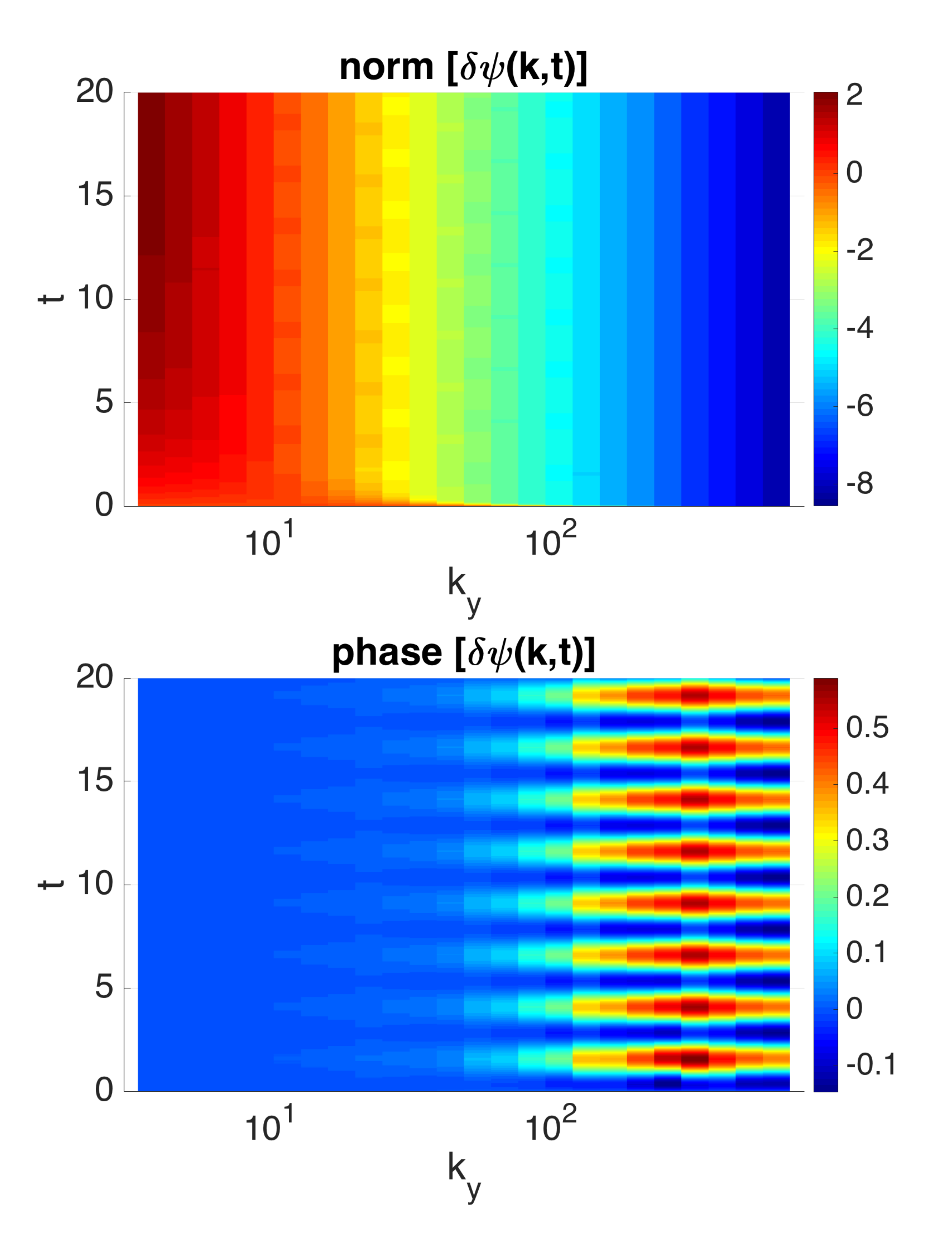}
\caption{\label{fig11} (top) The logarithm of norm and (bottom) the phase of $\delta\psi_k$ as functions of mode number $k$, and time $t$. Here, $\beta=100$, $\epsilon=0$, and $F=G=1$. Similar results are found for the case with $F=G=10$.}
\end{figure} 
\begin{figure}
\includegraphics[width=4cm, height=6cm]{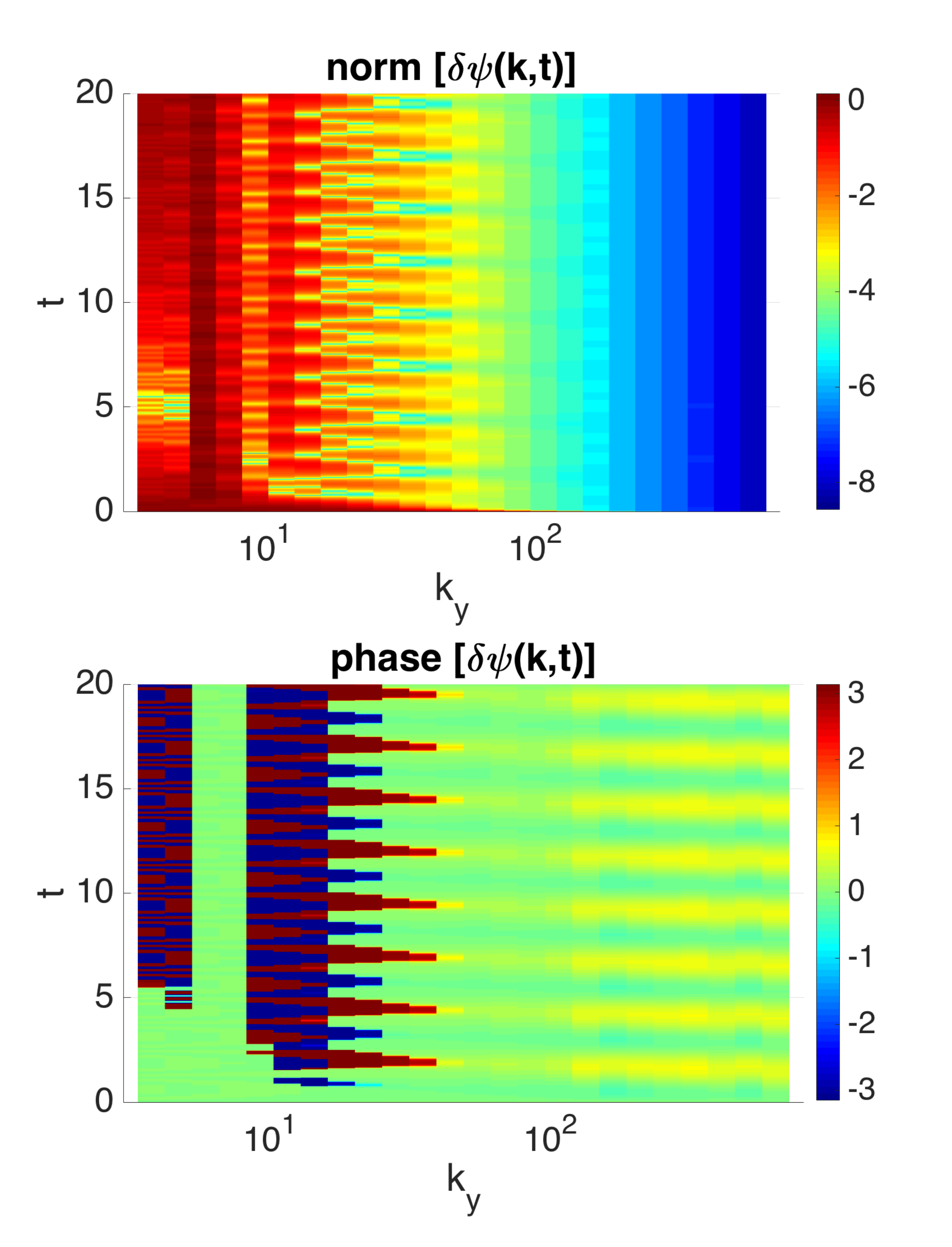}
\includegraphics[width=4cm, height=6cm]{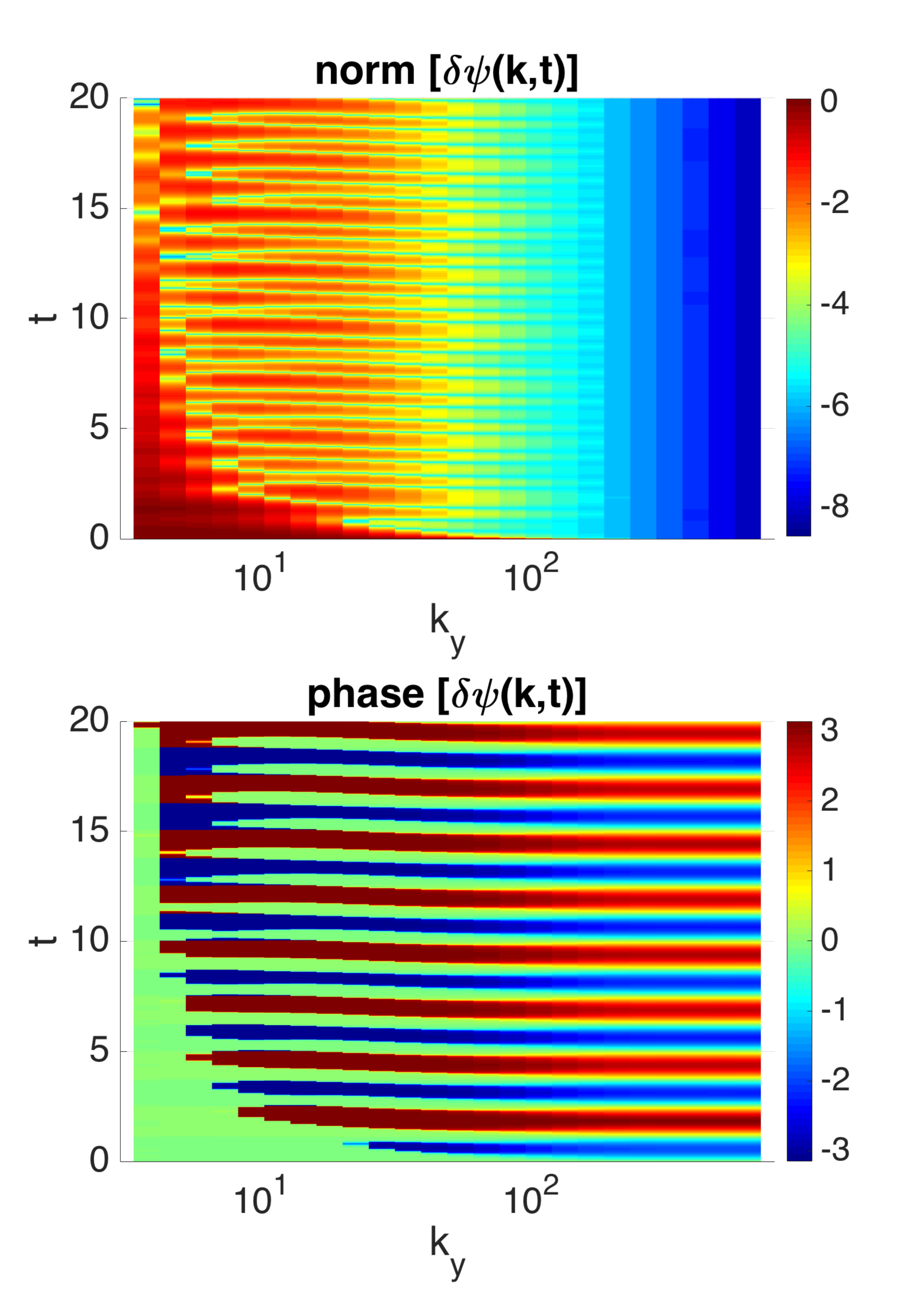}
\caption{\label{fig12} (top) The logarithm of norm and (bottom) the phase of $\delta\psi_k$ as functions of mode number $k$, and time $t$. Left figures present the results for the $F=G=1$ and right figures present the computed values for $F=G=10$. Here, $\beta=100$ and $\epsilon=5$.}
\end{figure} 

\subsection{The case with the linear cross-coupling} 
In this section we include the linear cross-coupling term between the flow and the forcing by setting $\epsilon=5$. This value is chosen following the results shown in Figs. \ref{fig6} (b-d) where we observe the competition between the effects of $\beta$ and $\epsilon$ to be strong. Such cross-coupling will introduce a cross-correlation between the phases of the two random fields which in Ref. \cite{krommes2000b} was shown to play an important role in reducing the saturation level of the fluctuations in stochastic models of passive advection. However there, the cross-correlation was introduced via a statistical dependence while here it is introduced through a linear cross-coupling between the phases. Figures \ref{fig10} (a and b) illustrate the time evolution of $C(t)$ for different phase states and increasing values of $\beta$. When $\beta =0$ the fluctuation auto-correlation increases in time similarly in both synchronised and a-synchronised cases i.e. $F=G=1$, and $F=G=10$, respectively. The time evolution of the norm and phase of $\delta\psi_k$ shown in Figs. \ref{fig11}. Here we observe a separation between the high and low $k$ where in the low $k$, the phases sit at $0$ but the phases at high $k$ oscillate in an small angle between $0$ and $0.5$ with requilar frequency.

Similarly to the case without the linear cross-coupling the increase in $\beta$ results in a strong reaction of the saturation levels in both $F=G=1$, and $F=G=10$ states, and as the increase in $\beta$ desynchronises the phases of the forcing the saturation levels converge to the same value for both states. Small modulations are observed in the evolution of $C(t)$ when $F=G=10$ which decreases in amplitude as $\beta$ is increased. 

The evolution of norm and the phase of $\delta \psi_k$ for $\beta=100$ and $\epsilon=5 $ are presented in Figs. \ref{fig12} left (a-synchronised) and right (synchronised). The effect of high $\beta$ is to increase the frequency and the range of oscillation of the phases in a wide range in $k$, while the norm is strongly affected at low $k$ as it decays in time.

In summary, our results show that a linear cross-coupling between the phases of the random flow and the forcing, plays an important role in enhancement of the saturation levels of the fluctuations auto-correlation $C(t)$. We find that cross-coupling is most effective in synchronising the phases at high $k$, and as a result increasing the $C(t)$. The interplay between the desynchronisation of the dispersion relation and synchronisation effect of the linear cross-coupling plays an important role in determining the steady state values of $C(t)$.

\section{Discussion} 
In this work, we have introduced a non-linear phase coupling model into the simplest model of the passive advection-diffusion of a scalar with forcing. The phase-coupling follows the well-established Kuramoto paradigm that has been shown to represent systems displaying self-organisation well. The model is intended to isolate the importance of the collective phase a-synchronisation/synchronisation states on the time evolution of the fluctuation energy and the possibility of the formation of coherent modulation structures. Such structures are observed in magnetically confined fusion plasma experiments where the self-organisation of DW turbulence forms the ZF which play an important role in the suppression of the DW turbulence itself allowing the access to high confinement modes of operation \cite{hillesheim2016}. The advective flow and the forcing are assumed as limit-cycle oscillators with a phase coupling model based on an extended version of the Kuramoto model. Here, the natural frequencies of the flow oscillators are assumed to follow a DW type dispersion relation, thus introducing a scale dependence in the phase coupling model. Our results show that the assumption of a fully stochastic phase state of the turbulence is more relevant for high values of scale separation with the energy spectrum following a $k^{-7/2}$ decay rate, while for lower scale dependence the a-synchronised and synchronised phases differ significantly, and one could expect the formation of coherent modulations in the latter case. 

\section{Acknowledgments}
We thank Professors P. H. Diamond, B. Knaepen and Ozg\"{u}r. D. G\"{u}rcan for very fruitful discussions. Sara Moradi has benefited from a mobility grant funded by the Belgian Federal Science Policy Office and the MSCA of the European Commission (FP7-PEOPLE-COFUND-2008 n$¼$ 246540).

\end{document}